# 面向 6G 通信感知一体化的固定与可移动天线技术

曾 勇[*1,2] 董珍君[1] 王蕙质[1] 朱立鹏[3]
洪子尧[1] 姜庆基[1] 王东明[1,2] 金 石[1] 张 瑞[3,4]
1. 东南大学移动通信全国重点实验室，江苏南京 210096；2. 紫金山实验室，江苏南京 211111；
3. 新加坡国立大学电子与计算机工程系，新加坡 117583；4. 香港中文大学（深圳），深圳大数据研究院，广东深圳 518172

**摘 要：**多天线技术通过在收发端部署天线阵列，从而提供额外的空间自由度（degrees of freedom，DoFs），大幅提升了无线通信的可靠性与有效性。与此同时，多天线技术应用于雷达感知领域，实现了空间角度分辨能力并提升了感知自由度，大幅增强了无线感知性能。然而，无线通信与雷达感知领域在过去数十年里独立发展。因此，尽管多天线技术在这两个领域分别取得了巨大的进步，但并没有通过发挥它们的协同作用来实现深度融合。随着感知与通信的融合被确定为第六代（the sixth-generation，6G）移动通信网络的典型应用场景之一，多天线技术的发展面临新的机遇以填补上述空白。为此，本文围绕未来天线阵列规模持续扩张、阵列架构更加多样、阵列形态更为灵活等发展趋势，对面向 6G 通信感知一体化的多天线技术进行综述。首先介绍未来多天线的不同架构类型，包括以传统紧凑式阵列和新兴稀疏阵列为代表的集中式阵列架构、以无蜂窝大规模 MIMO（multiple-input multiple-output）为代表的分布式天线架构，以及三维连续空间阵元位置与朝向灵活可调的可移动天线/流体天线。然后，本文将介绍基于上述天线架构的远场/近场信道建模，并进行通信与感知性能分析。最后总结不同天线架构的特点，并展望解决因天线阵列规模的持续扩展及阵列形态的灵活多变引起的信道状态信息获取困难的新思路。

# Fixed and Movable Antenna Technology for 6G Integrated Sensing and Communication

Zeng Yong[*1,2] Dong Zhenjun[1] Wang Huizhi[1] Zhu Lipeng[3]
Hong Ziyao[1] Jiang Qingji[1] Wang Dongming[1,2] Jin Shi[1] Zhang Rui[3,4]
1. National Mobile Communications Research Laboratory, Southeast University, Nanjing, Jiangsu 210096, China;
2. Purple Mountain Laboratories, Nanjing, Jiangsu 211111, China;
3. Department of Electrical and Computer Engineering, National University of Singapore, Singapore 117583, Singapore;
4. The Chinese University of Hong Kong (Shenzhen), Shenzhen Research Institute of Big Data, Shenzhen, Guangdong 518172, China

**Abstract:** By deploying antenna arrays at the transmitter/receiver to provide additional spatial-domain degrees of freedom (DoFs), multi-antenna technology greatly improves the reliability and efficiency of wireless communication. Meanwhile, the application of multi-antenna technology in the radar field has achieved spatial angle resolution and improved sensing DoF, thus significantly enhancing wireless sensing performance. However, wireless communication and radar sensing have undergone independent development over the past few decades. As a result, although multi-antenna technology has dramatically advanced in these two fields separately, it has not been deeply integrated by exploiting their synergy. A new opportunity to fill up this gap arises as the integration of sensing and communication has been identified as one of the typical usage scenarios of the sixth-generation (6G) mobile communication network. Motivated by the above, this article aims to explore the multi-antenna technology for 6G integrated sensing and communication (ISAC), with the focus on its future development trends such as continuous expansion of antenna array scale, more diverse array architectures, and more flexible antenna designs. First, we introduce several new and promising antenna architectures, including the centralized antenna architectures based on traditional compact arrays or emerging sparse arrays, the distributed antenna architectures exemplified by the cell-free massive multiple-input multiple-output (MIMO), and the movable/fluid antennas with flexible

---

*通讯作者：曾勇 yong_zeng@seu.edu.cn *Corresponding Author：Zeng Yong, yong_zeng@seu.edu.cn
基金项目：国家重点研发计划(2019YFB1803400)；国家自然科学基金面上项目(62071114)；国家自然科学基金(62331022)；广东省基础与应用基础研究重大专项(2023B0303000001)
Foundation Items: National Key R&D Program of China (2019YFB1803400); National Natural Science Foundation of China (62071114); National Natural Science Foundation of China (62331022); Guangdong Major Project of Basic and Applied Basic Research(2023B0303000001) positions and/or orientations in a given three-dimensional space. Next, for each antenna architecture mentioned above, we present the corresponding far-field/near-field channel models and analyze the communication and sensing performance. Finally, we summarize the characteristics of different antenna architectures and look forward to new ideas for solving the difficulties in acquiring channel state information caused by the continuous expansion of antenna array scale and flexible antenna designs.

## 1 引言

多天线技术通过在收发端引入天线阵列并有效设计各阵元的信号，从而在不额外消耗时频资源的情况下，大幅提升无线通信的可靠性和有效性，已发展成为现代无线通信系统的关键物理层技术之一。事实上，利用多天线进行无线信号传输具有悠久的发展历史[1]，最早可追溯至 1901 年马可尼的跨大西洋传输实验[2]。至 20 世纪 50 年代，相控阵技术开始发展[3]，通过调控发射端或接收端阵列各单元的信号相位，从而对目标信号方向形成波束，获得阵列增益。进一步，通过同时调控各阵列单元信号的相位和幅度，相控阵技术发展为自适应阵列，而后者进一步演进为智能天线，通过设计零陷对准干扰方向的阵列方向图，进一步实现干扰消除[4]。长期以来，无线通信系统中的多天线主要用于实现空间分集增益或阵列波束赋形增益，从而提高传输的可靠性。直至 20 世纪 90 年代，多输入多输出（multiple-input multiple-output，MIMO）技术的发展带来全新的空间复用增益[5-6]，即在同一时频资源并行发送多路数据流，从而提供了传统时频域维度以外的空间自由度，极大提升了频谱效率，推动通信系统迈入 MIMO 时代。早期 MIMO 系统一般使用 2~8 天线，通过空间复用和空间分集技术，显著提高了信号质量、抗干扰能力和系统容量，成功应用于无线局域网（IEEE 802.11）和第四代（the fourth-generation，4G）移动通信网络。2010 年提出的大规模 MIMO（massive MIMO）技术[7]，通过在基站端部署阵元数远大于用户数的大规模天线阵列，从而使得用户间的信道渐近正交并实现信道硬化，即小尺度衰落特性消失，信道趋于确定性。相较于传统 MIMO，大规模 MIMO 能极大提升系统频谱效率，已成为第五代（the fifth-generation，5G）移动通信系统的关键物理层技术。

此外，多天线技术在雷达感知领域也取得了长足发展，它克服了传统单天线雷达在目标检测、跟踪和识别上的局限，大幅提升感知性能。特别地，多天线技术的引入为雷达感知提供了测向能力，而其角度分辨率取决于阵列孔径。两种典型的多天线雷达模式包括 MIMO 雷达和相控阵雷[8]。对于 MIMO 雷达模式，发射阵列的各天线发送正交波形，从而形成对空间的全覆盖并获得波形分集增益，因而适用于无先验信息的目标搜索。而对于相控阵雷达，不同天线发送相干波形，从而对目标方向实现波束赋形获得相干处理增益，适用于有先验信息的目标跟踪。而对于目标搜索，由于相控阵雷达单波束周期内的感知区域受限，需进行周期性波束扫描才能实现对广域空间的感知覆盖。此外，对于包含 $M$ 发射阵元和 $N$ 接收阵元的单基地 MIMO 雷达，通过将发射端（或接收端）的阵元间距设置为接收阵列（或发射阵列）的总孔径，经过匹配滤波以后得到的感知信号输入输出关系可等效为具有 $MN$ 阵元的系统，即通过 $M+N$ 天线单元实现 $MN$ 维度的虚拟孔径，从而极大提高空间分辨率与感知自由度。

从天线规模而言，无线通信与雷达感知系统的技术演变遵循着相似的趋势，即从单天线到多天线，进一步拓展至大规模天线阵列。由于应用场景和性能指标的不同，无线通信与雷达感知系统长期以来独立发展。具体而言，在通信系统中，多天线通过分集增益实现高可靠传输，通过空间复用增益实现高速率传输。此外，大规模天线阵列的部署能够提升空间分辨率，极大增强了用户间干扰抑制能力以及高密度连接能力。而对于感知系统，多天线技术实现了角度分辨能力，提高了感知的空间自由度[9]，即可分辨目标的数量。多天线技术为无线通信与雷达感知系统都提供了分集增益、复用增益和空间分辨率。近年来，大型规模天线阵列和毫米波/太赫兹波段信号的组合使用，使得无线通信和雷达感知系统在硬件架构、信道特性和信号处理方法方面日趋相似，两者可以共同设计以高效利用资源，实现性能互惠增强，因此催生了面向通信和感知系统融合的多天线技术研究[10-21]。早期的研究主要聚焦于通信和感知系统的共存[10-13]，即通信和感知分别占用相互正交的空间资源，以抑制二者之间的干扰。例如，通信和感知系统采用不同的天线子阵，以实现空间资源正交[10]。或者，在通信的过程中，利用不同方向的波束实现额外的感知功能[11]。此外，还可以基于零空间投影的方式，将感知信号通过波束赋形的方法投影到通信信号的零空间内，以避免两者之间

的干扰[12-13]。然而，这类面向通感共存的方法，尽管能同时实现通信和感知功能，其资源利用率较低。因此，面向感知通信双功能的研究逐步兴起[14-18]，具体可以分为以感知为中心和以通信为中心两种设计思路。以感知为中心的方法，主要面向雷达感知系统以提供额外的通信功能。这类方法主要基于传统雷达系统，利用其空间波束的旁瓣调制通信符号[14-15]，或者基于天线选择进行指数调制[16]。因此，其通信性能（如通信频谱效率等）受限。另一方面，以通信为中心的方法主要以正交频率复用（orthogonal frequency division multiplexing, OFDM）为主[17-18]。这类方法基于额外的感知算法，在不显著改变通信系统的前提下，实现感知功能。然而，基于 OFDM 的感知面临高峰均功率比和高速移动场景下感知性能较差的问题。此外，为了充分挖掘通信感知一体化的性能，面向通信感知一体化的多天线技术也受到了广泛地关注[19-21]。然而，这类研究主要聚焦于传统紧凑式均匀天线阵列，通信和感知性能受限。

2023 年国际电信联盟无线电通信部门在《IMT 面向 2030 及未来发展的框架和总体目标建议书》中，将通信感知一体化确定为第六代（the sixth-generation，6G）移动通信网络的六大场景之一[22]。当前的通信感知融合系统中的多天线技术主要基于传统的紧凑式均匀天线阵列，对实现未来 6G 高性能通信感知仍具有挑战。因此 6G 通感一体化对多天线技术的进一步发展带来新的机遇与挑战，具体而言, 面向 6G 通信感知一体化的多天线技术面临以下发展趋势：

**阵列规模持续扩张**：无线通信与雷达感知的分集增益、复用增益、空间分辨率等性能上限受制于天线阵列规模或孔径。为支撑 6G 网络设想的宏伟愿景，如沉浸式通信、超高连接密度、超可靠性和低延迟以及超高感知定位精度[23-25]，未来基站的阵列规模将进一步扩张，有望从当前的 64 天线提升至 512 甚至更多，由大规模 MIMO 向超大规模 MIMO（extremely large-scale MIMO，XL-MIMO）演进[26-30]。然而，随着小区半径的进一步缩小及近距离感知需求的进一步提升，面向通信感知一体化的 XL-MIMO 并非是阵列规模的简单增加，而是带来了由传统远场均匀平面波到近场非均匀球面波的转变，为通信及感知的信道建模、性能分析、优化设计等方面带来了新的机遇与挑战[30]。特别地，对于无线通信系统，阵列规模的持续扩张有望提供超高空间分辨率，不仅使得用户间信道渐近正交，将进一步使得同一用户的多径信道在空间上正交，从而为解决多径符号间干扰提供了新的思路。通过挖掘这一特点，文献[31-32]提出了时延多普勒对齐调制（delay-doppler alignment modulation, DDAM）方法，通过逐径波束赋形及时延多普勒补偿，有效降低信道的时延多普勒扩展，极大提升了单载波及多载波传输的效率并降低信号的峰均比。

**阵列架构更加多样**：典型的多天线阵列将阵元以半波长间距规则地进行排列，称之为紧凑式阵列。未来多天线阵列架构的演进呈现了多样化的发展趋势。一方面，阵元间距可进一步减小直至趋于零，从而实现连续表面的全息 MIMO 或大型智能表面（large intelligent surface，LIS）[33]。不同于传统基于阵列响应向量信道模型的离散天线架构，连续表面需要考虑基于格林函数或傅里叶变换的连续空间电磁信道模型。另一方面，为了以较小阵元数实现更大的阵列孔径，显而易见的方法是增大阵元间距，使得阵元间距不再受半波长的约束，称为稀疏阵列[34]。事实上，稀疏阵列在无线定位及雷达感知领域的潜力已被充分挖掘，其通过对阵元间的信号进行（共轭）相关，形成更大维度的虚拟天线阵列，从而提升空间分辨率。然而，基于信号（共轭）相关实现大维虚拟阵列的方法难以直接应用于通信系统，其本质原因是通信系统需要解码通信符号，而对不同天线的信号直接进行（共轭）相关会导致用户间通信符号的非线性混叠。然而，已有研究初步表明[35-36]，即使不形成虚拟阵列，得益于天线孔径的增加，相同阵元数的稀疏阵列能实现比传统紧凑式阵列更佳的性能，特别是在用户密集场景。上述紧凑式阵列和稀疏阵列的天线单元都部署于同一平台，阵列的结构化特征明显，因此二者都属于集中式天线阵列。通过进一步增加天线阵元间的空间跨度，我们可以得到分布式天线阵列。

**阵列形态更为灵活**：传统天线阵列设计部署以后，阵元位置即进行固化。固定式天线由于天线孔径及间距限制，多个天线单元在空间呈离散化排布，使得多天线系统无法充分挖掘和利用无线信道在空间的连续变化，并难以在三维空间灵活排布。因此，固定式天线限制了通信系统的空间自由度，在信道深度衰落的情况下通信性能受限。可移动天线[37-38]或称流体天线[39-40]等可以有效解决固定式天线的性能瓶颈问题，具体而言，可移动天线能够在三维连续空间灵活改变位置和朝向，从而改善通信系统的信道条件并提升通信性能。相比于固定式天线，可移动天线能够获得完全的空间分集增益、空间复用增益以及灵活波束赋形，在提升通信速率及可靠性等方面具有较大潜力[37]。

围绕上述多天线技术发展新趋势，本文拟对面向 6G 通信感知一体化的多天线技术进行综述。首先介绍未来超大规模天线阵列的架构类型，包括紧凑式天线阵列、稀疏天线阵列、分布式天线阵列、及可移动

天线，进而不同天线架构的建模与通信感知性能分析，最后进行总结与展望。

## 2　天线阵列架构类型

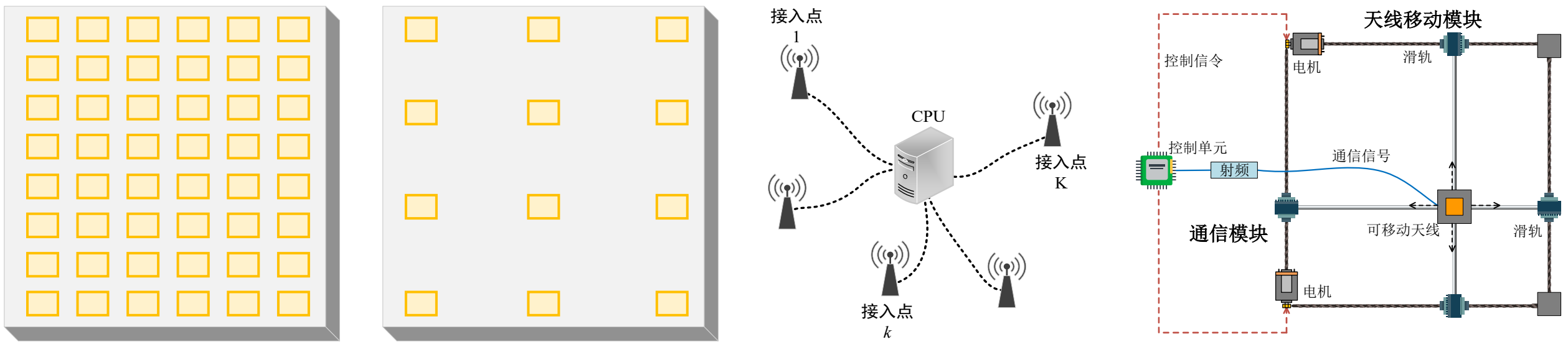

(a) 紧凑式天线阵列：$d = \lambda / 2$　(b) 稀疏天线阵列：$d = \eta\lambda / 2,\ \eta > 1$　(c) 分布式天线阵列　(d) 可移动天线

图 1　不同天线阵列架构

Fig. 1　Different antenna array architectures

根据天线阵列单元是否可移动，天线架构可以分为传统固定式天线和未来可移动天线，其中固定式天线架构根据阵列的部署结构，可分为集中式和分布式天线架构，而集中式天线架构根据相邻天线阵元是否大于半波长，分为紧凑式天线架构和稀疏天线架构。不同天线架构如图 1 所示，具体描述如下所述。

**紧凑式天线阵列**：如图 1(a)所示典型的紧凑式天线阵列架构，其所有天线阵元部署于同一平台上，阵元间距为 $d = \lambda / 2$，其中 $\lambda$ 表示信号波长。半波长间隔的紧凑式天线阵列可以避免角度模糊，其阵列的工作频率决定了相邻阵元的物理间距。鉴于半波长阵元间隔的约束，为提升阵列孔径及空间分辨率，紧凑式天线阵列仅能通过增加阵元数目来实现，通常伴随更高的硬件成本、更复杂的信号处理，以及更苛刻的部署平台要求。

**稀疏天线阵列**：除了像紧凑式天线阵列通过增加阵元数提高天线孔径，也可像稀疏天线阵列通过增大阵元间距提高天线孔径，如图 1(b)所示。根据任意相邻阵元的间距是否相等，可以将稀疏阵列进一步分为均匀与非均匀稀疏阵列。均匀稀疏阵列定义为阵元间距为 $d = \eta\lambda/2$ 的均匀阵列[35]，其中 $\eta > 1$ 定义为阵列的稀疏度。在阵元数量相同的情况下，均匀稀疏阵列相比于传统的紧凑式天线阵列具有更大的孔径，因而可以实现更窄的主瓣，获得更高的空间分辨率与干扰抑制能力。但同时，由于阵元间距的增大而导致的栅瓣效应将带来角度模糊问题[35-36]，为通信系统的用户间干扰抑制及感知系统的角度分辨带来新的挑战。然而，对于感知系统，通过利用非均匀稀疏阵列，对阵元位置进行优化设计并对阵元信号进行（共轭）相关等处理构造虚拟和差阵列，能够利用少量的物理天线实现较大虚拟孔径，因而在降低硬件成本的同时避免栅瓣。典型的非均匀稀疏阵列包括最小冗余阵列[41]、嵌套阵列[9]、互质阵列[42]和模块化阵列[36]。在阵元数相同的情况下，最小冗余阵列具有最大的空间自由度，但不同于嵌套阵列和互质阵列，最小冗余阵列的阵元位置只能通过穷举法确定，没有闭式解。对于无线通信系统，模块化天线阵列是一种新的有前景架构[36]，可以在实现比紧凑式天线阵列更高分辨率的同时，在一定程度上抑制稀疏阵列导致的栅瓣问题。模块化天线的阵元以模块化的方式规则地排列在同一部署平台，每个模块由适度规模的均匀天线阵列组成，相邻模块的间距可以根据部署平台的结构灵活确定。比如，模块化天线阵列可以部署于建筑物表面，模块之间由不适合部署天线的窗户等分隔，实现与建筑物表面的共形。

**分布式天线阵列**：不同于集中式天线阵列，分布式天线阵列是由一个大地理区域上的多个分布式接入点（access point，AP）组成，AP 通过光纤或无线信道与控制系统的中央处理器（central processing unit，CPU）连接，如图 1 (c)所示。用户不需由所有天线提供服务，而仅由附近的一组分布式天线集合来提供。通过 CPU 进行节点间的联合处理，可以提供宏分集增益和广域更均匀的覆盖，有效降低干扰。典型的分布式天线系统包括协调多点[43]、云无线电接入网络[44]、网络 MIMO[45]和无小区大规模 MIMO[46]等。然而，分布式天线架构也面临着新的挑战：(1) 同步要求更为严苛：由于射频器件的非理想性，不同 AP 的时钟难以完全一致。在通信系统中，同步只需要抑制终端与 AP 间的时钟偏差和信号传播时延的综合效应，从而保障通信的可靠性与有效性。而对于感知系统，由于其依赖于精确的信号传播时延估计以获得目标的位置与速度，同步要求更为严苛。目前通信设备 10 ns 级的同步误差会引入米级的定位误差，相干处理时长内

的时钟偏差与频率偏移漂移，及上下行切换引起的时域随机相位变化，对精确度要求较高的感知任务均构成显著挑战；(2) 额外下行导频开销：由于分布式天线阵列信道硬化效果相对集中式阵列更弱，用户更难以估计准确的等效信道增益，需要额外的下行导频来辅助用户估计准确的等效信道增益；(3) 前传容量受限：分布式天线系统受制于前传链路的容量，需要考虑分布式处理单元的信息交换量，以在容量限制和时延下完成协同处理。

**可移动天线**：不同于传统固定式天线，可移动天线能够在三维连续空间灵活改变阵元位置和朝向，为多天线无线通信与感知系统提供了新的设计自由度, 如图 1(d)所示。基于机械滑台的可移动天线系统可靠性强、机械结构相对稳定，可以支持不同类型的天线移动，并且天线定位精度高，但是电机和滑台结构通常尺寸较大，相应的硬件成本和功耗也较高。相比之下，流体天线结构尺寸相对较小、功耗较低，更适用于小型通信设备。但是流体天线由于液态属性，通常只能放在一维的流体管道内，且天线定位精度较低，辐射效率相比于传统固体天线更低且可靠性低。上述两种可移动天线结构在硬件实现上相对简单，但都面临机械移动速度受限的问题。相比之下，基于微机电系统的可移动天线[47]能够实现微秒级甚至纳秒级的响应速度和亚微米级的控制精度，能够实现更高效、更灵活的天线位置部署。此外，像素天线[48]作为一种新兴的天线技术，可以通过控制大量微型像素单元的连接来改变天线的位置和性质，相比于机械式移动天线具有更高的响应速度和更低的功耗。可移动天线并不局限于上述几种天线结构，而是强调利用天线的空间位置及朝向灵活性改善通信及感知性能。高效率、低能耗、灵活可控的可移动天线仍在持续研究当中。

表 1 总结了不同天线架构的优缺点和在实际部署中面临的挑战。现有的天线架构的天线阵元位置是固定的，即包含集中式天线和分布式天线的传统固定式天线。而未来可移动天线，由于天线单元的可移动性，因此与现有天线架构不兼容。在现实部署过程中，由于集中式天线需要在同一平台部署天线单元，因此面临苛刻的部署平台要求，而其中紧凑式天线相比于稀疏天线的部署要求更严格。传统紧凑式天线，其优势在于部署简单，但其结构单一且固化，在实现面向未来大孔径高分辨率通信感知一体化时，其硬件成本往往较高。因此，其通常适合部署在基站、高楼、山顶等，以提供广域稳定的通信和感知覆盖。对于稀疏天线，其优势在于部署灵活，能以小规模阵元实现大孔径高分辨率，但其问题在于存在栅瓣。因此，针对其可灵活部署的特点，其适合部署在建筑物表面、隧道，桥梁等不连续表面上，以提供额外的通信和感知覆盖。分布式天线因为天线单元部署在不同的地理位置，因此可以解决集中式天线部署难题。对于分布式天线，其优点在于可提供宏分集增益，覆盖面广，但其对不同站点的同步要求高。因此，可以利用小区间基站的定时和同步网络进行联合部署，以提供跨小区的通信和感知无缝切换服务。而在新兴的可移动天线技术中，实际部署的难易取决于设计的可移动平台的复杂性。对于可移动天线，其优点在于结构灵活，具有丰富的空间设计自由度，但其对天线移动位置的精确度要求高，且频繁移动开销大。为此，可移动天线可部署在如卫星网络、车联网、无人机网中，利用其载体本身的移动性和可控性，为特定的通信和感知一体化场景定制服务。为实现相同的阵列孔径，相比较于集中式天线阵列，稀疏天线阵列、分布式天线阵列以及可移动天线所需天线阵元数更少。在相同阵列孔径下，稀疏天线阵列、分布式天线阵列以及可移动天线的成本与能耗取决于各自具体的天线布置。

**表 1　不同天线架构特点**

**Tab. 1　Characteristics of different antenna array architectures**

| 天线架构 | | | 天线可移动性 | 部署结构 | 优点 | 缺点 |
|---|---|---|---|---|---|---|
| 固定式天线 | 集中式天线 | 紧凑式天线 | 固定 | 同一平台，需连续 | 结构简单、无栅瓣 | 结构固化、大孔径高分辨率增加硬件成本 |
| | | 稀疏天线 | 固定 | 同一平台，可非连续 | 以小规模阵元实现大孔径、高分辨率、可灵活部署 | 存在栅瓣 |
| | 分布式天线 | | 固定 | 不同分布式站点 | 提供宏分集增益、覆盖面积更大、有效降低干扰 | 同步要求严苛、额外下行导频开销、前传容量限制 |

| | | | | |
|---|---|---|---|---|
| 可移动天线 | 可移动 | 三维连续空间 | 结构灵活、提供新的自由度、高空间分集与复用增益、灵活波束赋形 | 移动速度受限、频繁移动开销大 |

# 3　集中式天线阵列通信与感知

本节介绍适用于紧凑式和稀疏天线的集中式架构通信与感知系统建模和性能分析。首先讨论紧凑式和稀疏阵列架构的远场信道建模和性能分析，然后拓展至考虑非均匀球面波的近场通信与感知。

## 3.1　集中式天线阵列远场建模与性能分析

### 3.1.1　远场建模

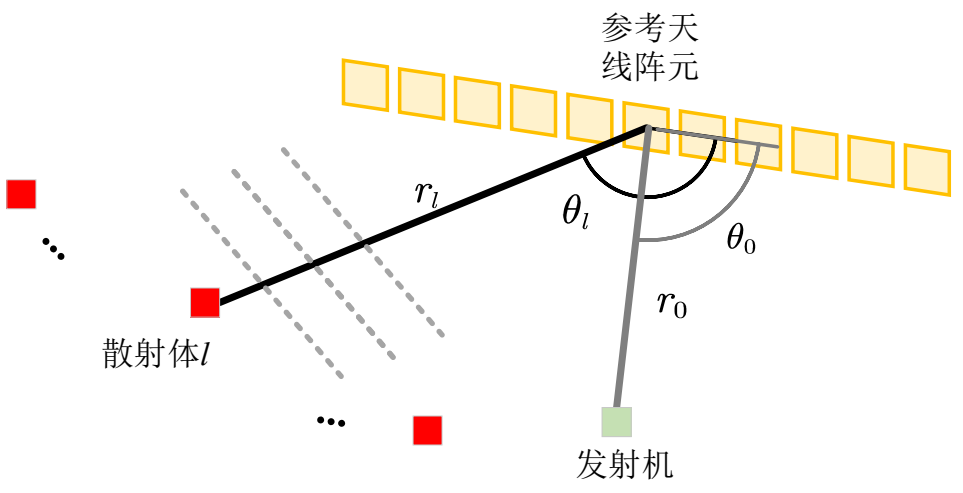

图 2　集中式天线阵列远场信道模型

Fig. 2　Far-field channel model for centralized antenna arrays

如图 2 所示，考虑基站端配置 $M$ 阵元的均匀线性阵列。以线阵的中心阵元作为参考阵元，远场阵列响应向量可表示为

$$\boldsymbol{a}^{\mathrm{UPW}}(\theta)=\left[\exp\left(-\mathrm{j}2\pi/\lambda\cdot md\cos\theta\right)\right]_{m\in\mathcal{M}},\quad \mathcal{M}=\{-(M-1)/2,\ldots,(M-1)/2\} \tag{1}$$

其中 $\theta$ 为信源的角度，$d$ 为相邻阵元间隔，以及 $\mathcal{M}$ 为天线索引集合。对典型紧凑式均匀阵列，$d=\lambda/2$，而对于稀疏阵列，$d=\eta\lambda/2$，其中 $\eta>1$ 为稀疏度。远场阵列响应向量只与信源角度 $\theta$ 相关，与信源距离无关，且阵元间的相位是线性变化的。基于阵列响应向量，收发端之间的远场信道向量 $\boldsymbol{h}\in\mathbb{C}^{M\times 1}$ 建模为

$$\boldsymbol{h}=\sqrt{\frac{K_c}{K_c+1}}\alpha_0\boldsymbol{a}^{\mathrm{UPW}}(\theta_0)+\sqrt{\frac{1}{K_c+1}}\sum_{l=1}^{L}\alpha_l\boldsymbol{a}^{\mathrm{UPW}}(\theta_l) \tag{2}$$

其中 $K_c$ 为莱斯因子，$\theta_0$ 为视距链路的信号方向，$L$ 为多径数量，$\theta_l$ 和 $\alpha_l$ 分别为第 $l$ 个径的信号方向和归一化增益。由于考虑远场模型，上述建模表现出两个主要特点：首先，同一路径的信号在不同阵元的信号方向是相同的；其次，同一路径信号在不同阵元的复增益也是相同的。因此，上述模型中，$\theta_l$ 和 $\alpha_l$ 均与天线索引无关。当用户端配置 $N$ 阵元的均匀阵列，MIMO 远场信道矩阵 $\boldsymbol{H}\in\mathbb{C}^{M\times N}$ 可建模为

$$\boldsymbol{H}=\sqrt{\frac{K_c}{K_c+1}}\alpha_0\boldsymbol{a}_{\mathrm{r}}^{\mathrm{UPW}}(\theta_{\mathrm{r},0})\left(\boldsymbol{a}_{\mathrm{t}}^{\mathrm{UPW}}(\theta_{\mathrm{t},0})\right)^{\mathrm{H}}+\sqrt{\frac{1}{K_c+1}}\sum_{l=1}^{L}\alpha_l\boldsymbol{a}_{\mathrm{r}}^{\mathrm{UPW}}(\theta_{\mathrm{r},l})\left(\boldsymbol{a}_{\mathrm{t}}^{\mathrm{UPW}}(\theta_{\mathrm{t},l})\right)^{\mathrm{H}} \tag{3}$$

其中 $\boldsymbol{a}_{\mathrm{r}}^{\mathrm{UPW}}(\theta_{\mathrm{r},l})=\left[\exp(-\mathrm{j}2\pi/\lambda\cdot md\cos\theta_{\mathrm{r},l})\right]_{m\in\mathcal{M}}$ 和 $\boldsymbol{a}_{\mathrm{t}}^{\mathrm{UPW}}(\theta_{\mathrm{t},l})=\left[\exp(-\mathrm{j}2\pi/\lambda\cdot nd\cos\theta_{\mathrm{t},l})\right]_{n\in\mathcal{N}}$ 分别为接收端和发送端的远场阵列响应向量，$\mathcal{N}=\{-(N-1)/2,\ldots,(N-1)/2\}$ 为发送天线索引集合，$\theta_{\mathrm{r},l}$ 和 $\theta_{\mathrm{t},l}$ 分别为多径信号的到达角和离开角。

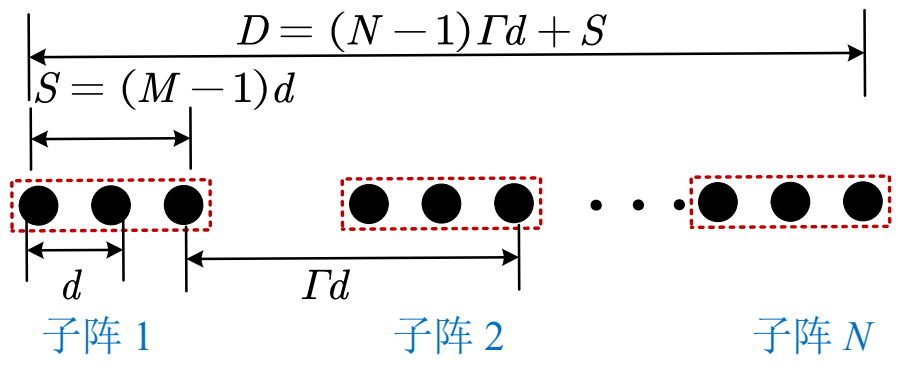

图 3　模块化天线阵列

Fig. 3　Modular antenna array

远场阵列响应向量的具体表达形式取决于天线架构。特别地，对于模块化天线阵列，如图 3 所示，由 $N$ 个 $M$ 阵元的紧凑式均匀线阵模块构成，相邻模块间距离为 $\Gamma d$，其中 $\Gamma$ 是依赖于实际部署结构的模块化

参数，且 $d=\lambda/2$ 。因此，每个模块的物理尺寸为 $S=(M-1)d$ ，总物理尺寸为 $D=(N-1)\Gamma d+S$ 。将式(1)的远场阵列响应向量应用于不同的集中式线性阵列，可得紧凑式与稀疏阵列响应向量表达式，总结如表 2。特别地，表 2 中模块化天线阵列的远场阵列响应向量可以表示为一个阵元间距为 $\Gamma d$ 的 $N$ 阵元稀疏阵列与一个 $M$ 阵元的紧凑式均匀线阵的阵列响应向量的克罗内克乘积的形式[36]。

**表 2　不同集中式线阵架构的远场阵列响应向量**

**Tab. 2　Far-field array response vectors for different centralized linear array architectures**

| 天线架构 | | 远场阵列响应向量 |
|---|---|---|
| 紧凑式阵列 | | 公式(1)： $d=\lambda/2$ |
| 稀疏阵列 | 均匀稀疏阵列 | 公式(1)： $d=\eta\lambda/2$ |
| | 模块化阵列 | $\boldsymbol{a}^{\mathrm{UPW}}(\theta)=\boldsymbol{p}(\theta)\otimes\boldsymbol{b}(\theta)$<br>$\boldsymbol{p}(\theta)=\left[\exp\left(-\mathrm{j}2\pi/\lambda\cdot n\Gamma d\cos\theta\right)\right]_{n\in\mathcal{N}},\boldsymbol{b}(\theta)=\left[\exp\left(-\mathrm{j}2\pi/\lambda\cdot md\cos\theta\right)\right]_{m\in\mathcal{M}}$ |

### 3.1.2　远场通信感知一体化

多天线阵列远场波束图的主瓣宽度定义为波束图零点之间的间隔，其与具体的天线架构密切相关。表 3 总结了不同线阵架构的波束图闭式表达式、主瓣宽度和栅瓣位置。为公平比较，不同天线架构的阵元数皆设为 $NM$。远场情况下不同架构的波束图仅与空间频率差 $\Delta_\theta=\cos\theta-\cos\theta'$ 相关，主瓣宽度随着阵列稀疏度的增大而减小。特别地，模块化天线阵列的远场波束图可表示为一个 $N$ 阵元间距为 $\Gamma$ 个半波长的均匀稀疏阵列的波束图与 $M$ 元紧凑式线阵的波束图的乘积[36]。

**表 3　不同集中式线阵架构的波束图、主瓣宽带及栅瓣位置**

**Tab. 3　Beam pattern, main-lobe beam width and grating lobes for different centralized linear array architectures**

| 天线架构 | | 波束图 | 主瓣宽度 | 栅瓣位置 |
|---|---|---|---|---|
| 紧凑式阵列 | | $G_{MN}^{\mathrm{co}}(\theta,\theta')=\left\lvert\dfrac{\sin\left(0.5\pi MN\Delta_\theta\right)}{MN\sin\left(0.5\pi\Delta_\theta\right)}\right\rvert$ | $BW=\dfrac{4}{MN}$ | 无 |
| 稀疏阵列 | 均匀稀疏阵列 | $G_{MN}^{\mathrm{spa}}(\theta,\theta')=\left\lvert\dfrac{\sin\left(0.5\pi MN\eta\Delta_\theta\right)}{MN\sin\left(0.5\pi\eta\Delta_\theta\right)}\right\rvert$ | $BW=\dfrac{4}{MN\eta}$ | $\Delta_\theta=2n/\eta$<br>$n=\pm1,\ldots,\pm\lfloor\eta\rfloor$ |
| | 模块化阵列 | $G_{MN}^{\mathrm{mod}}(\theta,\theta')=\left\lvert\dfrac{\sin\left(0.5\pi N\Gamma\Delta_\theta\right)}{N\sin\left(0.5\pi\Gamma\Delta_\theta\right)}\right\rvert\left\lvert\dfrac{\sin\left(0.5\pi M\Delta_\theta\right)}{M\sin\left(0.5\pi\Delta_\theta\right)}\right\rvert$ | $BW=\dfrac{4}{N\Gamma}$ | $\Delta_\theta=2n/\Gamma$<br>$n=\pm1,\ldots,\pm\lfloor\Gamma\rfloor$ |

图 4 比较了不同集中式阵列架构的远场波束图，其中阵元数设为 $NM$=16，模块化天线阵列模块数设为 $N$=4 个模块天线阵元数 $M$=4 块间距参数为 $\Gamma=13$，均匀稀疏阵列的稀疏度设为 $\eta=2.8$ 。由图 4 可见，相较于传统紧凑式阵列，均匀稀疏阵列和模块化阵列都具有更好的角度分辨率。然而，由于阵元间距大于半波长，两类稀疏阵列均产生栅瓣效应，对通信与感知系统优化设计提出了新的挑战。

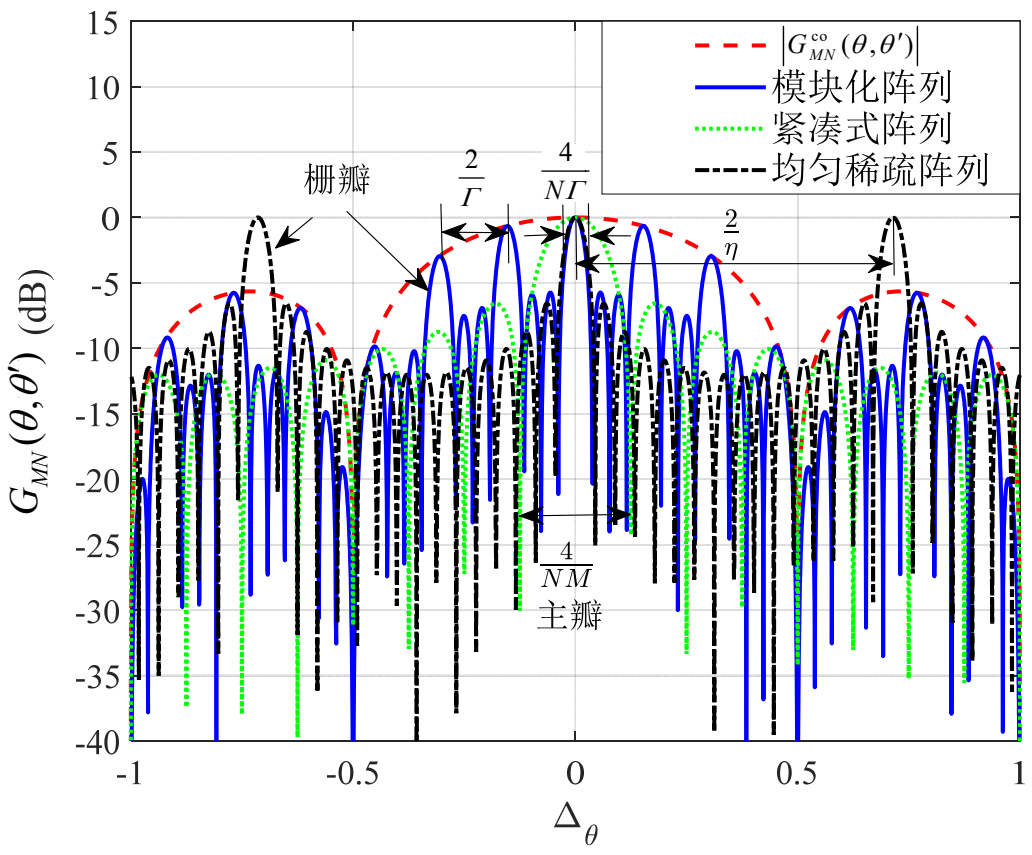

图 4　不同集中式线阵架构的波束图[36]
Fig. 4　Beam pattern for different centralized linear array architectures[36]

基于以上远场波束图分析，相较于传统紧凑式天线阵列，在相同阵元数情况下，稀疏阵列由于具有更窄的主瓣宽度，有望降低用户密集区域的用户间干扰。然而，其产生的栅瓣效应将使得位于栅瓣位置上的用户间干扰增大。因此，传统紧凑式天线阵列与稀疏阵列的抗干扰性能比较需要考虑上述因素的折中关系。文献[35]展示了假设任意用户角度独立均匀分布于 $\theta \in [-\theta_{\max}, \theta_{\max}]$，任意两个用户之间的空间频率差 $\Delta_\theta$ 并非均匀分布，而是任意两个用户空间频率差的概率密度随着 $|\Delta_\theta|$ 增大而降低。这实际上为稀疏天线阵列的栅瓣提供了天然的抑制，即任意两用户处于对方栅瓣的概率要小于其处于主瓣的概率。通过挖掘这一新特性，文献[35]通过理论推导证明了在用户密集区域，均匀稀疏阵列能够取得比传统紧凑式阵列更好的通信性能。图 5 比较了均匀稀疏阵列和紧凑阵列的用户通信速率的累计误差分布函数，其中阵元数为 32，均匀稀疏阵列的稀疏度为 $\eta = 4$ 。观察到在 90%百分位，均匀稀疏阵列的通信速率是紧凑式阵列的四倍[35]。

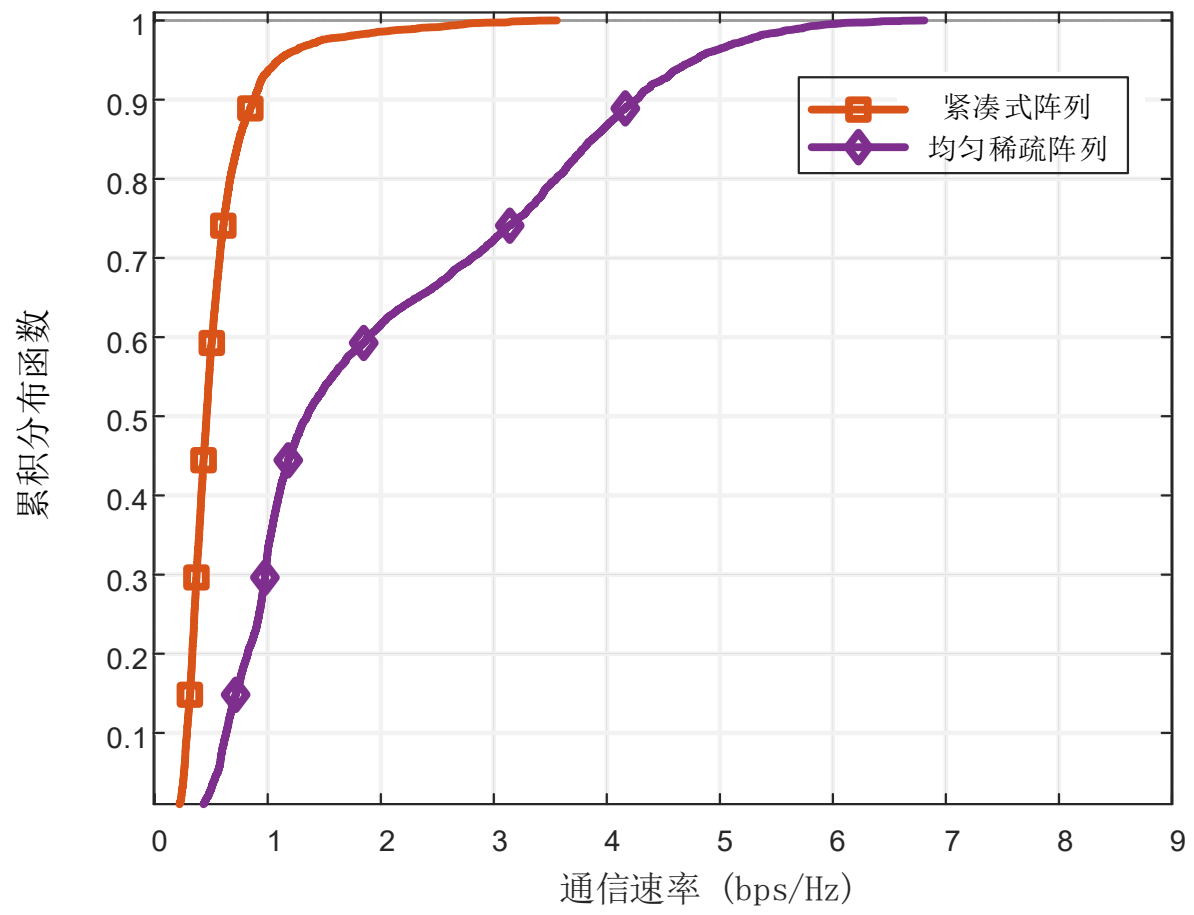

图 5　紧凑式阵列与均匀稀疏阵列通信速率的累计分布[35]
Fig. 5　The cumulative distribution function (CDF) of data rate for uniform sparse array and compact array[35]

当前通感一体化研究大多考虑传统紧凑式天线阵列，在波形设计、信号处理等多个方面进行了深入的研究[17][49]。具体地，针对不同的感知任务与传输波形，其问题建模与信号处理流程通常不同。以通信领域常见的 OFDM 波形为例，单站 OFDM 通感一体化系统可以用于感知目标的时延 $\tau = 2R/c$ 和多普勒 $f_d = 2v/\lambda$，其中 $R$ 为目标与雷达的距离，$c$ 为光速，$v$ 为目标的径向移动速度。则由空间中 $K$ 个目标反射的回波信号可以表示为

$$y(t) = \sum_{k=1}^{K} \gamma_k x(t-\tau_k)\exp(\mathrm{j}2\pi f_{d,k} t) + \omega(t) \tag{4}$$

其中 $x(t)$ 为发送的基带 OFDM 信号，$\gamma_k$ 为目标回波的幅度，$\omega(t)$ 为加性高斯白噪声，$\tau_k$ 和 $f_{d,k}$ 分别是第 $k$ 个目标的时延与多普勒参数。经过去除循环前缀并消除已知通信信号的影响等信号处理流程，得到用于感知的信号为

$$\boldsymbol{F} = \boldsymbol{A}_{\mathrm{D}}^{\mathrm{RX}} \mathrm{diag}(\boldsymbol{\gamma})(\boldsymbol{A}_{\mathrm{R}}^{\mathrm{RX}})^{\mathrm{T}} + \boldsymbol{\Omega} \tag{5}$$

其中 $\boldsymbol{\gamma} = [\gamma_1, \gamma_2, \cdots, \gamma_K]^{\mathrm{T}}$ 为 $K$ 个用户的目标回波幅度，$\boldsymbol{A}_{\mathrm{D}}^{\mathrm{RX}}$ 和 $\boldsymbol{A}_{\mathrm{R}}^{\mathrm{RX}}$ 分别对应时延与多普勒参数的等效转向矩阵，$\boldsymbol{\Omega}$ 是噪声矩阵。由于 $\boldsymbol{A}_{\mathrm{D}}^{\mathrm{RX}}$ 和 $\boldsymbol{A}_{\mathrm{R}}^{\mathrm{RX}}$ 的形式与传统角度估计问题中的转向矩阵相似，因而可以通过采用传统角度估计算法，例如 MUSIC、ESPRIT 等子空间类算法实现距离以及多普勒的超分辨估计。

此外，通过对无线通信信道环境的感知、识别与预测，有望进一步提升无线通信系统的性能。在传统通信系统中，通常需要采用穷搜的方法实现发射与接收端的波束匹配，导致大量的导频开销与时延。通过感知的先验位置信息能够大幅降低波束扫描的范围，进而降低实时扫描开销[50]。文献[51]提出了一种利用超分辨感知实现极低导频开销的信道估计方案，极大提高了通信质量。当通信链路建立之后，收发双方需要依赖实时校准确保通信质量，通过集成在发射机的感知单元回传最优波束索引，可大幅度降低这一部分

的开销。此外，对于高速移动的场景，可利用感知信号实现波束预测估计高动态性的通信信道。

稀疏天线阵列在通信感知一体化领域的研究仍然处于起步阶段，但其发展前景非常广阔。目前大多数运行在 6GHz 以下频段的蜂窝网络（如 4G 和 5G）由于测距和角度分辨率较低，只能提供米级精度的感知，难以满足智能车联网等应用需求[52]。面向未来 6G 的超密集连接需求，传统的紧凑式天线阵列难以支撑超高通信速率与超高感知精度的需求，而稀疏阵列将有望以其超高的空间分辨率，实现密集分布场景下的高性能通信感知一体化。文献[53]研究了一种由稀疏智能超表面辅助的通信感知一体化系统，通过在智能超表面上应用少量稀疏分布的主动传感器来进行有效的信道估计，从而实现针对通信和感知目标的优化波束成形，仿真结果验证了稀疏阵列辅助的通信感知一体化的有效性，并比较了不同稀疏阵列配置的性能。

## 3.2 集中式天线阵列近场建模与性能分析

### 3.2.1 近场建模

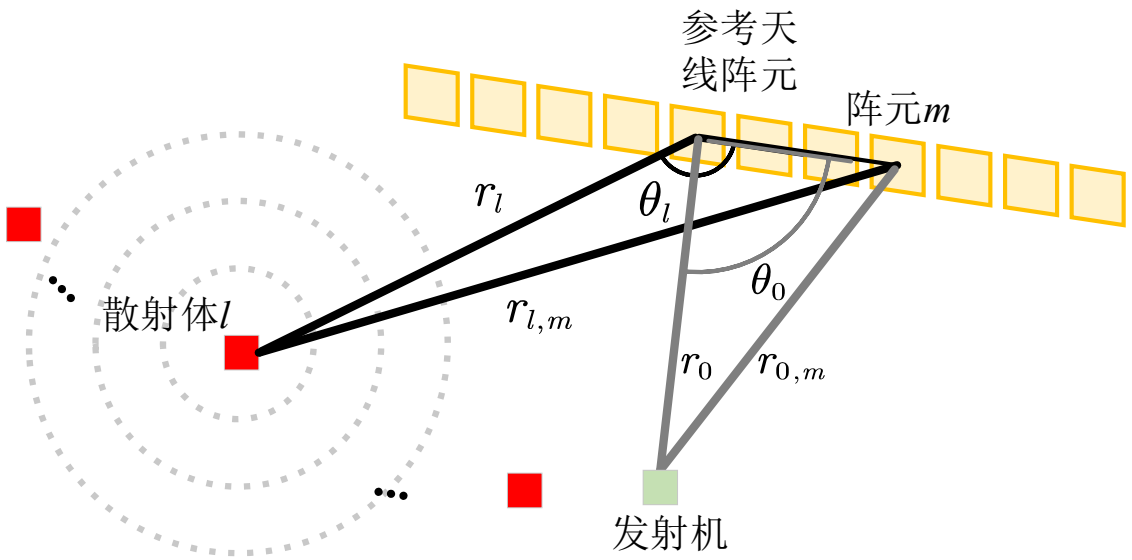

图 6 集中式天线阵列近场建模

Fig. 6 Near-field channel model for centralized antenna arrays

如图 6 所示，当 XL-MIMO 阵列的尺寸增大，信源与阵列中心的距离和阵列尺寸相当时，阵元之间的信号强度变化不可忽略，且阵元间的相位变化不再是线性的，因此需要考虑更一般的近场非均匀球面波模型。近场阵列响应向量表示为[29]

$$\boldsymbol{a}^{\mathrm{NUSW}}(r,\theta)=[r/r_m\cdot\exp(-\mathrm{j}2\pi/\lambda(r_m-r))]_{m\in\mathcal{M}} \tag{6}$$

其中 $r$ 为信号源到参考阵元的距离，$r_m=\sqrt{r^2+2rmd\cos\theta+(md)^2}$ 为第 $m$ 个阵元到参考阵元的距离，其不能像远场模型一样通过一阶泰勒近似。对比式(1)和(6)可见，近场阵列响应向量不仅和信源的角度$\theta$有关，还和距离 $r$ 相关。近场阵列响应向量的近场效应表现在两个方面：首先，相位是通过精确距离来表示，而不是它的一阶泰勒近似，因此不同天线阵元间的相位变化是非线性的；其次，不同阵元的信号强度也有所差异，体现在式(6)中的 $r/r_m$ 项。基于非均匀球面波模型，收发端之间的近场信道向量可建模为

$$\boldsymbol{h}=\sqrt{\frac{K_c}{K_c+1}}\alpha_0\boldsymbol{a}^{\mathrm{NUSW}}(r_0,\theta_0)+\sqrt{\frac{1}{K_c+1}}\sum_{l=1}^{L}\alpha_l\boldsymbol{a}^{\mathrm{NUSW}}(r_l,\theta_l) \tag{7}$$

其中 $K_c$ 为对应于参考阵元的莱斯因子，$\boldsymbol{a}^{\mathrm{NUSW}}(r_0,\theta_0)$ 和 $\boldsymbol{a}^{\mathrm{NUSW}}(r_l,\theta_l)$ 分别为视距链路和多径 $l$ 的近场阵列响应向量，$r_l$ 与 $\theta_l$ 分别为多径 $l$ 所对应的散射体与参考阵元之间的距离及角度。当用户端也配置 $N$ 阵元的阵列时，近场 MIMO 信道矩阵 $\boldsymbol{H}\in\mathbb{C}^{M\times N}$ 可建模为

$$\boldsymbol{H}=\sqrt{\frac{K_c}{K_c+1}}\boldsymbol{a}_{\mathrm{r}}^{\mathrm{NUSW}}(r_{\mathrm{r},0},\theta_{\mathrm{r},0})\left(\boldsymbol{a}_{\mathrm{t}}^{\mathrm{NUSW}}(r_{\mathrm{t},0},\theta_{\mathrm{t},0})\right)^{\mathrm{H}}+\sqrt{\frac{1}{K_c+1}}\sum_{l=1}^{L}\alpha_l\boldsymbol{a}_{\mathrm{r}}^{\mathrm{NUSW}}(r_{\mathrm{r},l},\theta_{\mathrm{r},l})\left(\boldsymbol{a}_{\mathrm{t}}^{\mathrm{NUSW}}(r_{\mathrm{t},l},\theta_{\mathrm{t},l})\right)^{\mathrm{H}} \tag{8}$$

其中 $\boldsymbol{a}_{\mathrm{r}}^{\mathrm{NUSW}}(r_{\mathrm{r},l},\theta_{\mathrm{r},l})=\left[\frac{r_{\mathrm{r},l}}{r_{\mathrm{r},l,m}}\exp\left(-\mathrm{j}\frac{2\pi}{\lambda}(r_{\mathrm{r},l,m}-r_{\mathrm{r},l})\right)\right]_{m\in\mathcal{M}}$ 和 $\boldsymbol{a}_{\mathrm{t}}^{\mathrm{NUSW}}(r_{\mathrm{t},l},\theta_{\mathrm{t},l})=\left[\frac{r_{\mathrm{t},l}}{r_{\mathrm{t},l,n}}\exp\left(-\mathrm{j}\frac{2\pi}{\lambda}(r_{\mathrm{t},l,n}-r_{\mathrm{t},l})\right)\right]_{n\in\mathcal{N}}$ 分别为接收端和发送端的近场阵列响应向量。上述基于收发端阵列响应向量外积的近场模型，使得近场信道矩阵 $\boldsymbol{H}$ 的秩的上限为 $L$+1，具有一定的局限性。比如，纯视距链路情况下，rank($\boldsymbol{H}$)=1，不适用于近距离场景。为此，文献[29]给出了直接考虑收发阵列对幅度和相位关系的更一般化近场 MIMO 模型。

近场阵列响应向量的具体表达式取决于阵列架构，如表 4 所示。特别地，对于模块化天线阵列，为简化近场阵列响应向量建模，只考虑相位在阵元上非线性变化而忽略幅度变化，可以进一步区分两种场景[36]：

(1) 子阵非共角，即不同模块 $n$ 的信号角度 $\theta_n$ 不同。此时，链路距离 $r$ 满足 $2S^2/\lambda \le r < 2D^2/\lambda$，其中 $S$ 和 $D$ 分别为每个模块和整个阵列的尺寸。这时，信源位于每个阵列模块的远场区域内，但在整个阵列的近场区域内。因此，每个模块 $n$ 内可采用均匀平面波模型，而不同模块间采用球面波模型；(2) 子阵共角，即不同模块的角度 $\theta_n$ 可近似相等，即 $\theta_n \approx \theta$， $\forall n$，此时链路距离相较于子阵非共角场景更大，满足 $\max\{5D, 4SD/\lambda\} \le r < 2D^2/\lambda$，其阵列响应可进一步简化，如表 4 所示。

**表 4 不同集中式线阵架构的近场阵列响应向量**

**Tab. 4 Near-field array response vectors for different centralized linear array architectures**

| 天线架构 | | 近场阵列响应向量 | |
|---|---|---|---|
| 紧凑式阵列 | | 如公式(6)： $d=\lambda/2$ | |
| 稀疏阵列 | 均匀稀疏阵列 | 如公式(6)： $d=\eta\lambda/2$ | |
| | 模块化阵列 | (1) 子阵非共角： $\boldsymbol{a}^{\mathrm{NUSW}}(r,\theta)=\left[\exp\left(-\mathrm{j}2\pi/\lambda(r_n-r)\right)\boldsymbol{b}(\theta_n)\right]_{n\in\mathcal{N}}$， $r_n$ 为第 $n$ 个模块的距离 | (2) 子阵共角： $\boldsymbol{a}^{\mathrm{NUSW}}(r,\theta)=\boldsymbol{e}(r,\theta)\otimes\boldsymbol{b}(\theta)$, $\boldsymbol{e}(r,\theta)=\left[\exp\left(-\mathrm{j}2\pi/\lambda(r_n-r)\right)\right]_{n\in\mathcal{N}}$ |

### 3.2.2 近场通信感知一体化

对于传统远场通信与感知，阵列响应向量只与角度相关，与距离无关，而基于球面波的近场系统同时具有角度和距离分辨率能力，有望实现更优的抗干扰及空间复用性能。通过采用高稀疏度的稀疏阵列增大阵列孔径，利用近场非均匀球面波模型增加空间自由度及分辨率，将有望进一步提高通信系统的容量。文献[36][54-56]针对模块化超大规模天线阵列，根据近场非均匀球面波模型，推导了近场信噪比闭合表达式，进一步揭示了其信噪比缩放定律和渐近性能，以及与传统远场均匀平面波模型结果的差异性。从图 7 可以看出，当模块数趋于无穷，相比于平面波模型，非均匀球面波模型下的信噪比趋于常数而非无限增长[54-55]，满足能量守恒定律。

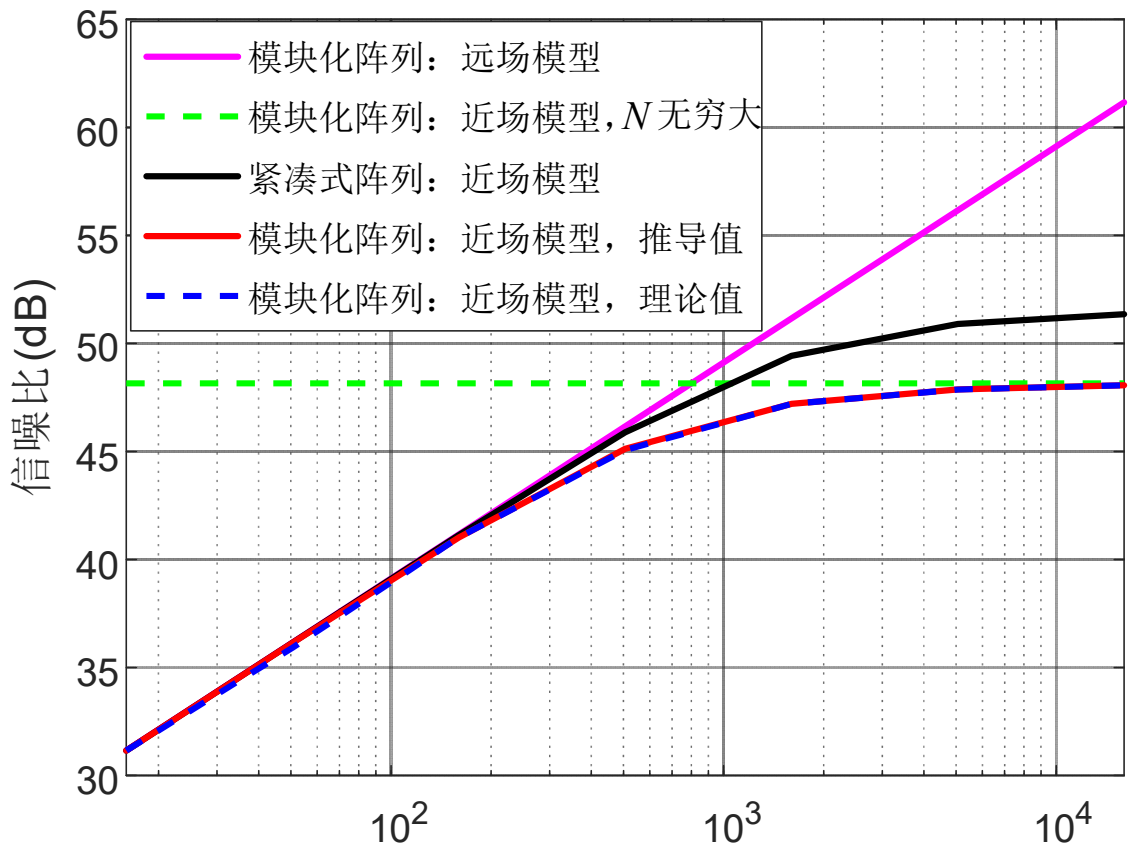

图 7 不同集中式线阵接收信噪比随天线数目变化关系[54]

Fig. 7 Signal-to-noise ratio (SNR) versus the number of array elements for different centralized linear arrays[54]

近场场景下不同架构的波束聚焦图不仅与空间频率差 $\Delta_\theta$ 有关，还和距离差 $\Delta_r = r - r'$ 相关。文献[56]针对模块化阵列结构特点，基于子阵非共角与子阵共角的球面波模型，分析了其近场波束聚焦图。基于子阵非共角均匀球面波模型，模块化超大规模线性阵列的近场波束聚焦图取决于从不同阵列模块中观察到的位置 $(r_n,\theta_n)$ 和 $(r_n',\theta_n')$。它等价于 $N$ 个波束图的加权和的形式，且每个波束图有不同的空间频率差 $\Delta_{\theta,n}$。基于子阵共角均匀球面波模型下的近场波束聚焦图可分为两部分，第一部分为具有 $N$ 个元素的稀疏线性阵列的均匀球面波波束聚焦图，第二部分为具有 $M$ 个元素的紧凑式线阵的均匀球面波波束聚焦图。图 8 所示紧凑式阵列和模块化阵列在近场信道模型下的波束聚焦图。从图 8 中观察到，与具有相同天线数量的紧凑式阵列结构相比，模块化天线阵列可以显著提高角度和距离维度的空间分辨率，但会产生栅瓣。

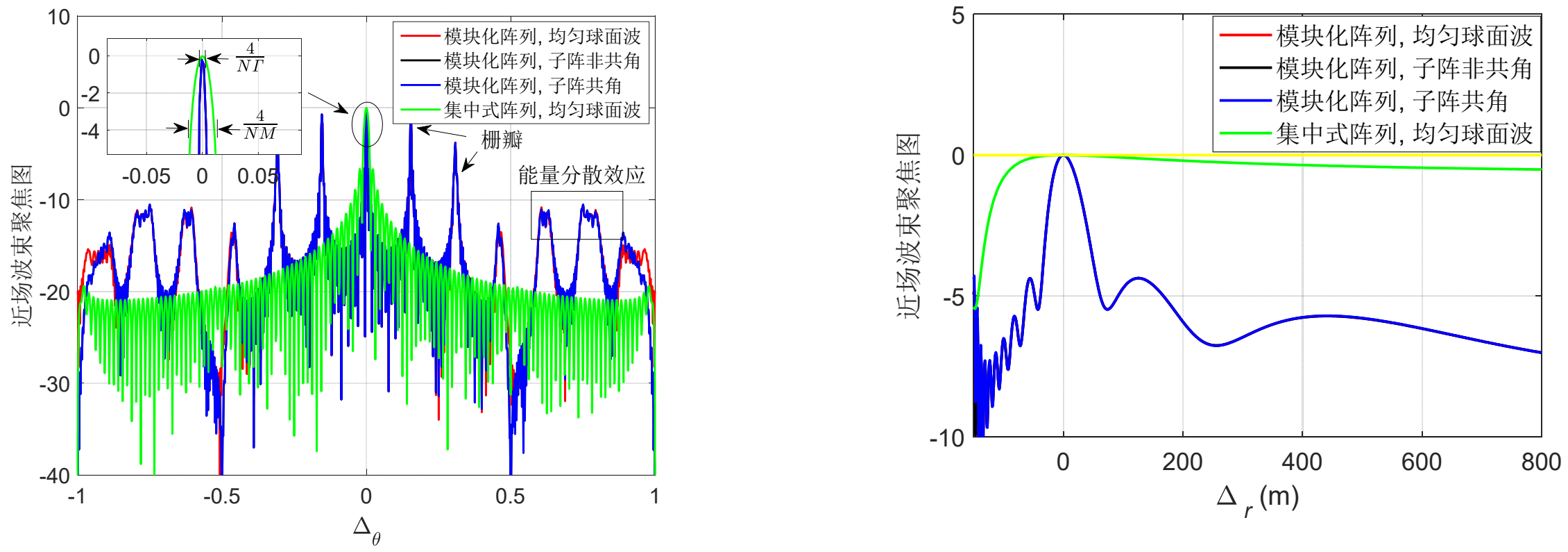

(a) 近场波束聚焦图随空间频率差 $\Delta_\theta$ 的变化关系　　(b) 近场波束聚焦图随距离差 $\Delta_r$ 的变化关系

图 8　基于近场信道模型的不同集中式线阵的波束聚焦图[56]

Fig. 8　The beam focusing patterns of different centralized linear arrays based on near-field channel model[56]

为了解决栅瓣问题，文献[36]针对近场多用户模块化天线阵列通信系统，提出了基于贪婪算法的用户分组策略，使得配对用户不落在栅瓣，大大减少了用户间干扰。图 9 中可以看出，在用户密集分布情况下，相比于紧凑式阵列结构，模块化天线阵列可显著提升通信性能。

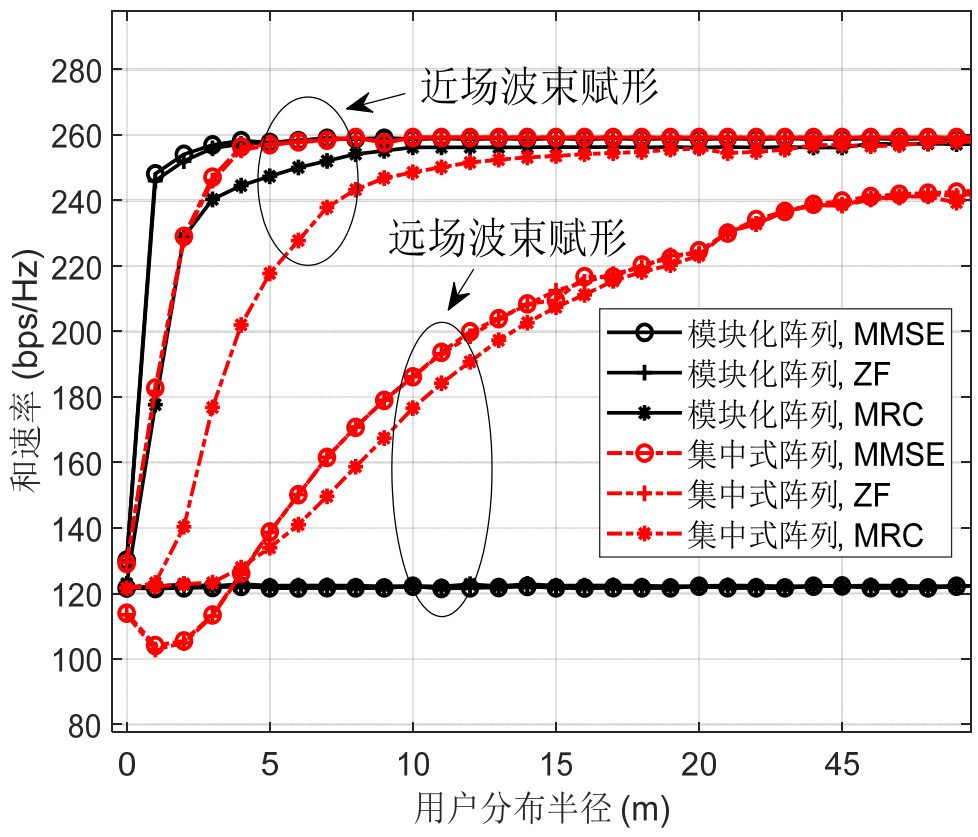

图 9　模块化和紧凑式阵列通信和速率随用户分布半径变化关系[36]

Fig. 9　Communication sum rates for modular and compact arrays versus the circular radius of user distribution[36]

除了在通信领域的性能增益，集中式天线阵列为近场感知带来了新的机遇与挑战。在远场感知系统中，角度和距离感知通常需要在空间域和频域分别进行处理，因而远场感知系统通常要求较大的带宽以及多节点协同实现精确定位。在近场感知系统中，回波信号中同时包含目标的角度和距离信息，因而单锚点、小带宽的精确定位成为了可能[57]，大大降低了系统的成本和复杂度。此外，远场感知系统只能估计目标的径向速度，无法获得目标的完整运动状态。而在近场感知中，不同的天线从不同角度“观察”感知目标，使得同时感知目标的径向速度和横向速度成为可能，进而进行精确的目标跟踪与预测。

此外，近场场景下的感知性能边界也呈现出与传统均匀平面波模型不同的变化趋势。基于二阶泰勒近似，文献[58]考虑了近场信号源定位和速度跟踪问题，并推导了后验克拉美罗下界。然而，尽管二阶泰勒近似方法在大多数实际系统中性能较好，其引入的系统误差将使模型不够精确[59]。此外，大多数现有的性能分析与定位算法仅针对单跳信号传播的信号源定位问题[60-61]，不能直接应用于具有双跳信号传播的雷达感知场景。文献[62]推导了近场感知的接收信噪比闭式表达式，不同于远场感知中的线性变化，近场场景下的信噪比随天线阵元数量呈现非线性变化，并最终趋于收敛。文献[63]考虑超大规模 MIMO 雷达与相控阵雷达，分别推导了近场场景下单站与双站感知系统角度以及距离参数的克拉美罗界的闭式表达式。文献[64]考虑了基于 MIMO 雷达三维近场定位的克拉美罗界及估计算法设计。图 10 展示了单站 MIMO 雷达与相控阵雷达近场感知的角度参数克拉美罗界与天线阵元数的变化关系。由此可见，随着天

线阵元数的不断增大，基于远场假设的克拉美罗下界将产生较大的误差。此外由于相控阵雷达存在波束赋形增益，因而超大规模相控阵雷达将获得更优的参数估计性能。

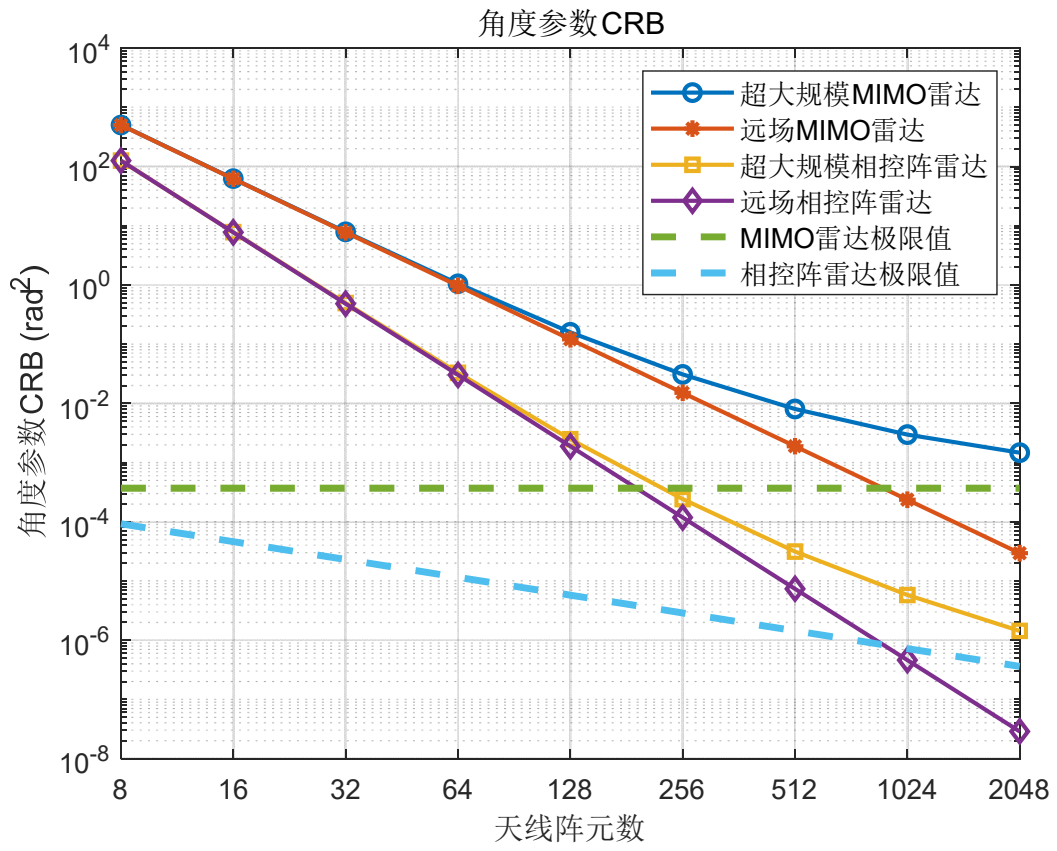

图 10 单站近场感知的角度克拉美罗界[63]

Fig. 10 Cramér–Rao Bound (CRB) of angle for monostatic near-field sensing[63]

虽然在近场感知的信号模型中，角度与距离参数高度耦合，因而能够更加准确地描述目标在空间中的位置。但同时，波达方向与距离的强耦合性导致大部分现有的超分辨率参数估计算法不再适用，很多学者直接将远场的感知算法进行扩展，实现了近场感知的角度、距离的联合估计。例如，通过将传统远场场景下的两种角度估计算法，即一维波束赋形算法与超分辨率 MUSIC 算法拓展至近场感知，实现角度距离参数的联合估计[65]。算法实现步骤与远场场景几乎一致，其差异主要存在于将一维角度域的谱峰搜索拓展为二维的谱峰搜索函数为：

$$
\begin{aligned}
P_{\text{2D-BF}}\left(r,\theta\right) &= \boldsymbol{a}^{\mathrm{H}}(r,\theta)\boldsymbol{R}_x\boldsymbol{a}(r,\theta) \\
P_{\text{2D-MUSIC}}\left(r,\theta\right) &= \frac{1}{\boldsymbol{a}^{\mathrm{H}}(r,\theta)\boldsymbol{E}_n\boldsymbol{E}_n^{\mathrm{H}}\boldsymbol{a}(r,\theta)}
\end{aligned}
\tag{9}
$$

其中 $\boldsymbol{R}_x$ 是接收信号矩阵的协方差矩阵，$\boldsymbol{E}_n$ 为将 $\boldsymbol{R}_x$ 经过 SVD 分解得到的噪声子空间。这类算法虽然估计精度较高，但是二维联合搜索使得算法的计算复杂度比传统远场信源定位要高得多。多维的近场谱峰搜索导致该算法的计算复杂度急剧增大。针对该问题，学者提出了很多改进算法，如多项式求根算法[66]、最大似然估计算法[67]、路径跟踪法[68]、加权线性预测法[69]等。近年来有学者提出将角度与距离参数解耦，从而将高维谱峰搜索过程转化为多个一维谱峰搜索，从而降低了近场信源定位算法的复杂度。首先，通过对 $r_m$ 进行二阶泰勒近似，将距离参数表示为 $r_m \approx r - md\sin\theta + m^2d^2\cos^2\theta/2r,\ \forall m \in \mathcal{M}$，并进一步将方向向量表示为

$$
\boldsymbol{a}(r_k,\theta_k) = \left[\exp\left(\mathrm{j}\phi_k m^2 + \mathrm{j}\omega_k m\right)\right]_{m\in\mathcal{M}}
\tag{10}
$$

其中 $\omega_k = -2\pi d\sin\theta_k/\lambda$ 和 $\phi_k = \pi d^2\cos^2\theta_k/(\lambda r_k)$ 分别是包含角度以及距离参数的中间变量。为了进一步解耦角度和距离参数，文献[70]将方向向量 $\boldsymbol{a}(r_k,\theta_k)$ 分解为

$$
\boldsymbol{a}(r_k,\theta_k) = \boldsymbol{F}\left(r_k,\theta_k\right)\xi\left(r_k,\theta_k\right)
\tag{11}
$$

其中 $\boldsymbol{F}\left(r_k,\theta_k\right) = \begin{bmatrix}\boldsymbol{P} \\ \boldsymbol{P}^{\mathrm{H}}[2:(M+1)/2]\end{bmatrix}$，$\boldsymbol{P} \triangleq \mathrm{diag}\left(\left[\exp(\mathrm{j}\omega_k m)\right]_{-(M-1)/2\le m\le 0}\right)$ 以及 $\boldsymbol{P}^{\mathrm{H}}[x:y]$ 表示取 $\boldsymbol{P}^{\mathrm{H}}$ 的第 $x$ 行至第 $y$ 行，$\xi\left(r_k,\theta_k\right) = \left[\exp(\mathrm{j}\phi_k m^2)\right]_{-(M-1)/2\le m\le 0}$。该算法能够实现将二维的角度-距离联合搜索问题转化为两个一维的谱峰搜索，从而在一定程度上降低复杂度。近年来，有学者提出并改进了基于二阶统计积量的算法[71-73]，即构建空间累积量为

$$
r(am+b,m) = \mathbb{E}\left(y_{am+b}(t)y_m^*(t)\right) = \sum_{k=1}^{K} p_k \exp\left(\mathrm{j}[(a-1)m+b]\omega_k + \mathrm{j}\left[(a^2-1)m^2 + 2abm + b^2\right]\phi_k\right)
\tag{12}
$$

其中 $y_m(t)$ 为阵元 $m$ 的接收信号，$p_k$ 为目标 $k$ 对应的反射系数。通过设计不同的参数 $a$ 和 $b$，可以消除 $\omega_k$ 或 $\phi_k$ 的影响，实现角度距离的解耦。由于需要多次矩阵分解操作，因此该类算法通常都需进行参数配对处理，但由于基于二阶统计量的算法不需要谱峰搜索，因而普遍具有计算复杂度低的特点。高阶统计量由于具有能够保持信号相位和去除高斯噪声干扰的良好特性，这使得基于高阶统计量的算法具有自动抑制加性高斯白噪声及色噪声的能力。虽然构造高阶统计量的近场感知算法不需要谱峰搜索，但由于需要构建高维度的累积量矩阵，计算的复杂度也相对较高，而如何有效地降低计算复杂度，避免谱峰搜索和参数配对，并且最大限度地提高参数估计的精度一直是近场源参数估计的关键点。另一方面，上述两类算法均要求阵元间距不大于四分之一波长，且存在无法准确估计同一方向不同距离的信源等局限性，限制了这些算法在实际中的应用。因此，高性能低复杂度的近场源定位算法仍需进一步系统且深入地研究。

近场场景下的通信与感知系统的建模与分析较为复杂，因而当前针对近场通感一体化系统的研究仍然处于起步阶段。早期的部分文献，分析与总结了近场通信感知一体化中潜在的机遇与挑战，探索了近场通信感知一体化的潜力[74-75]。进一步地，聚焦于近场通信感知一体化的波形设计或波束赋形的研究，文献[76-79]通过构造并求解优化问题实现近场场景下通信与感知性能的联合优化，针对近场球面波的不同建模方式，文献[76-77]考虑了均匀球面波模型，而文献[78-79]则基于更加精确的非均匀球面波模型进行优化分析。此外，上述近场通感一体化研究主要基于传统的紧凑式天线阵列，缺乏考虑稀疏阵列以及模块化天线阵列的近场通感一体化性能分析。

### 3.3 集中式天线通信与感知小结

相比于传统紧凑式天线阵列，稀疏阵列能够利用相同规模的天线数实现更大的阵列孔径，因而能够提高通信与感知的空间分辨率，在面向 6G 的超密集连接场景中具有更大的应用价值。同时，随着未来阵列稀疏度的进一步增大，其近场范围将大幅度增加，因而覆盖范围内的大部分通信与感知用户皆位于近场中。基于近场非均匀球面波传输特性，稀疏阵列的部署将不仅能够有效提高空间复用增益，进一步提高通信容量，还能够为感知系统提供新的距离与速度估计范式，有望实现超分辨率的定位与感知功能，缓解当前分布式定位模式下的同步与干扰难题。然而，稀疏阵列的栅瓣问题将更加突出，使得通信与感知面临较强的用户间干扰与定位模糊问题，通过合理选择稀疏度，例如部署模块化天线，将有望在一定程度上缓解栅瓣的影响。未来需要进一步研究栅瓣的抑制技术以及最大化通信与感知性能折中的阵列稀疏度优化方案等关键问题。

## 4 分布式天线通信与感知

### 4.1 系统建模

分布式天线的信道在 sub6G 频段通常被建模为莱斯或瑞利信道，而在毫米波频段信道通常采用多径信道模型，与紧凑式天线的信道建模本质上并没有不同，但建模的颗粒度在 AP 和用户层面上，整体的信道则是 AP 和用户信道的堆叠。显然，由于天线的大空间跨度分布，不同 AP 间的信道趋于独立，从而整体信道的相关度相较于集中式天线阵列降低，能够带来宏分集增益。考虑在 sub6G 频段下，第 $m$ 个单天线 AP 和单天线用户的信道 $h_m$ 可以表示为

$$h_m = \sqrt{\beta_m}\left(\sqrt{\frac{K_m}{K_m+1}}\exp\left(-\mathrm{j}2\pi/\lambda \cdot r_m\right) + \sqrt{\frac{1}{K_m+1}}g_m\right) \tag{13}$$

其中 $\beta_m$ 为第 $m$ 个 AP 对应的信道大尺度衰落，$K_m$ 为第 $m$ 个 AP 的莱斯因子，$r_m$ 是用户到第 $m$ 个 AP 的距离，$g_m \sim \mathcal{CN}(0,1)$ 为第 $m$ 个 AP 与用户间非视距链路信道，则用户的堆叠信道为 $\boldsymbol{h} = [h_1, h_2, \cdots, h_M]^{\mathrm{T}} \in \mathbb{C}^{M\times 1}$。当用户配置 $N$ 阵元的线阵时，第 $m$ 个 AP 与用户间的信道 $\boldsymbol{h}_m \in \mathbb{C}^{N\times 1}$ 可表示为

$$\boldsymbol{h}_m = \sqrt{\beta_m}\left(\sqrt{\frac{K_m}{K_m+1}}\exp\left(-\mathrm{j}2\pi/\lambda \cdot r_m\right)\boldsymbol{a}(\theta_m) + \sqrt{\frac{1}{K_m+1}}\boldsymbol{g}_m\right) \tag{14}$$

其中 $\boldsymbol{a}(\theta_m) = \left[\exp(-\mathrm{j}2\pi/\lambda \cdot nd\cos\theta_m)\right]_{n\in\mathcal{N}}$ 为第 $m$ 个 AP 与用户间视距链路的阵列响应向量，$\boldsymbol{g}_m \in \mathbb{C}^{N\times 1} \sim \mathcal{CN}\left(0, \boldsymbol{C}_m\right)$ 为第 $m$ 个 AP 与用户间非视距链路信道向量，$\boldsymbol{C}_m \in \mathbb{C}^{N\times N}$ 为第 $m$ 个 AP 和用户间信道的归一化协方差矩阵。因此，对于用户的堆叠 MIMO 信道矩阵表示为 $\boldsymbol{H} = [\boldsymbol{h}_1, \boldsymbol{h}_2, \cdots, \boldsymbol{h}_M]^{\mathrm{T}} \in \mathbb{C}^{M\times N}$。对于

毫米波分布式天线系统，第 $m$ 个 AP 和配置 $N$ 根天线的用户间的信道 $\boldsymbol{h}_m$ 可通过式(2)进行建模。

## 4.2 通信性能分析

对于分布式天线系统，其性能受具体处理方式的影响。根据信号处理的集中程度，可以将分布式天线系统分为四个层级：完全集中式的 Level 4 级、具有部分集中处理能力的 Level 3 级、完全分布式且不协作的 Level 2 级、完全分布式的小小区 Level 1 级[80]。具体来说，以上行为例，Level 4 是所有 AP 联合处理，这就需要将所有信息及信令汇总至 CPU，完全按照集中式天线的传统处理方式，显然会带来极高的前传代价；Level 3 则是先在本地仅使用本地信息进行处理，再在 CPU 进行合并，这时 CPU 需要一定的信息但远低于 Level 4；Level 2 则可看作 Level 3 的简化版，在 CPU 合并时不需要任何信息进行简单合并；Level 1 则是将 AP 看作一个微基站，每个 AP 最多只服务一个用户。而对于多层前传网络的分布式网络，可以实现 Level 4 和 Level 3/2 之间的折中，由边缘分布式单元实现一个集中式子网。如图 11 所示，在同一场景下，文献[80]对比了 Level 4 到 Level 1 以及宏基站的性能，展示了基于有效协作的分布式天线架构相较于传统蜂窝架构的性能增益。

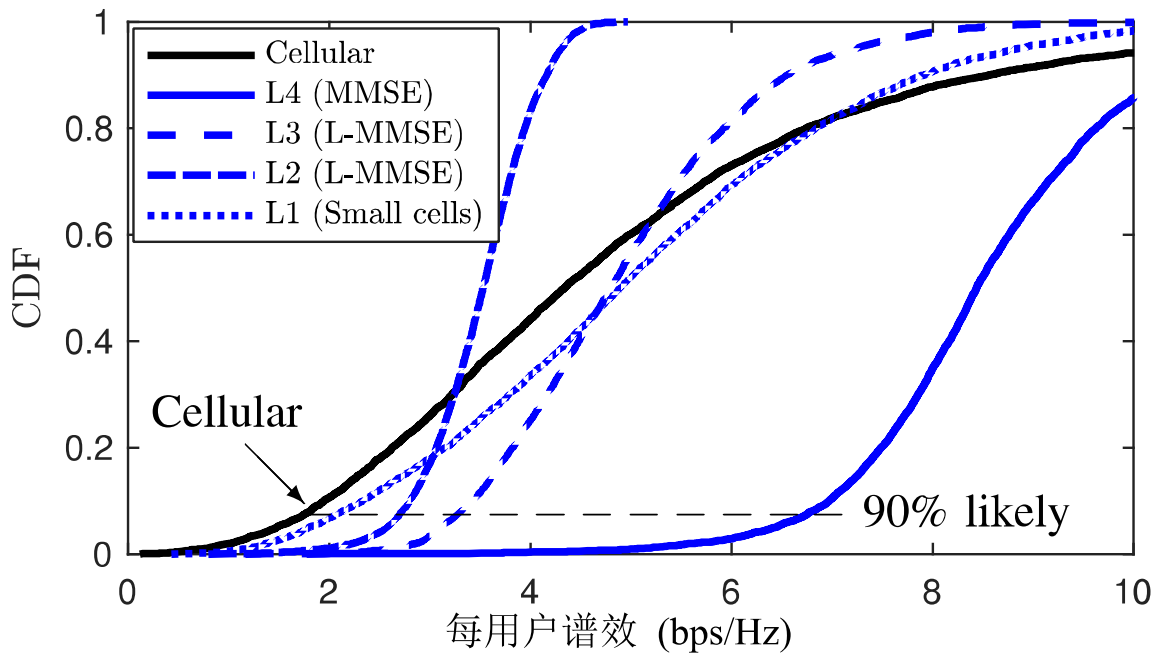

图 11 分布式网络的集中式和分布式处理结果对比[80]

Fig. 11 Comparison between centralized and distributed cell-free networks[80]

图 12 所示文献[81]中在实测场景下的层次前传结构的分布式天线的用户速率累积分布。在该系统下，采用了三层的前传网络结构，AP 和 CPU 之间插入了一层边缘分布式单元（edge distributed unit, EDU），形成一个三层的多叉树结构。在图 12 中 Local 最大比合并（maximal ratio combining，MRC）表示每个 AP 利用本地信道信息进行 MRC 接收，L-MMSE 表示 AP 进行本地最小均方误差（minimum mean squared error，MMSE）接收，P-MMSE, x EDUs 则表示采用了 x 个边缘分布式单元，每个边缘分布式单元进行本地的 MMSE 接收，最后 J-MMSE 表示完全集中的 MMSE 接收。可以看到合适的子网划分（如 P-MMSE, 4 EDUs）可以实现较高的性能和可接受的传输代价。

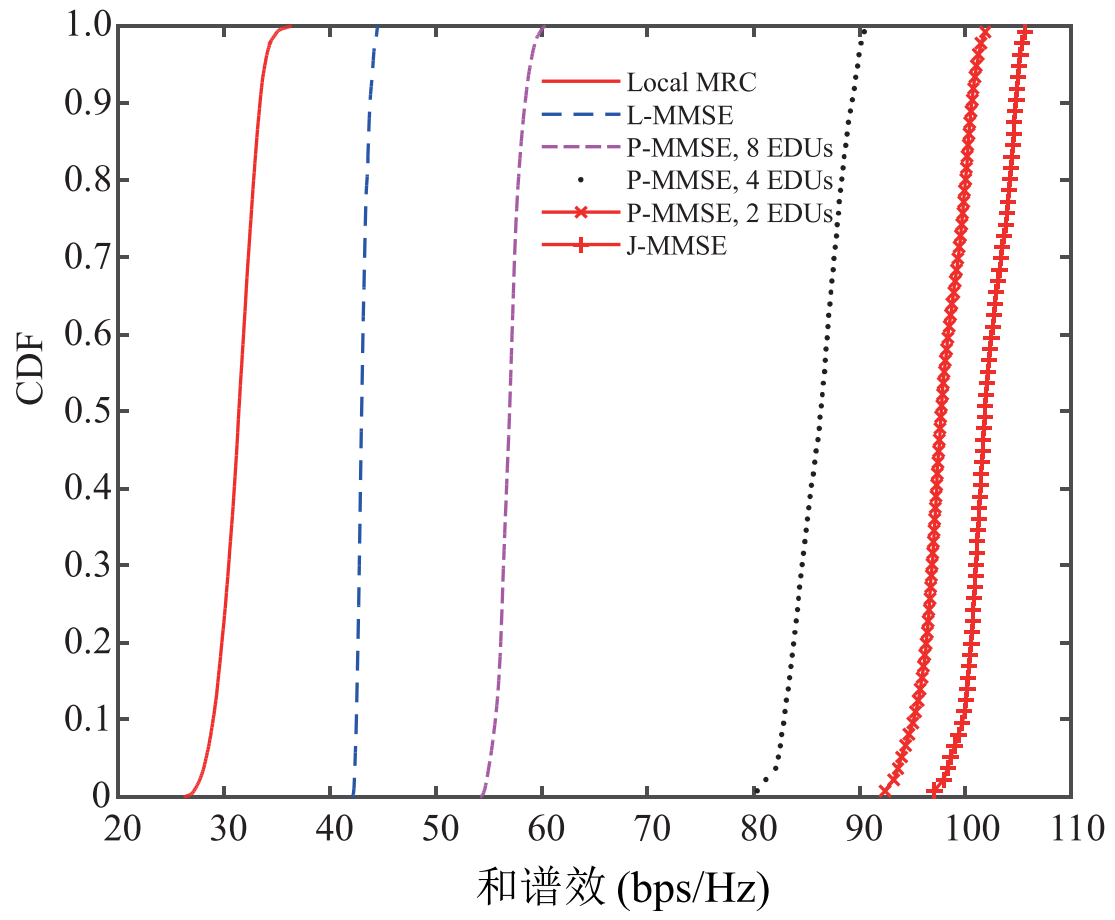

图 12 不同前传网络结构下的分布式网络性能对[81]

Fig. 12 Comparison between cell-free network with different fronthaul structures[81]

## 4.3 感知及通信感知一体化性能分析

基于分布式天线的感知系统研究，通常采用确定性建模，此时感知问题可以理解为参数估计问题。周

期图类算法[82]，可以基于 FFT 快速实现，但是算法性能有限且主要受限于系统带宽。子空间类算法[83]可以突破带宽限制，但是一般要求连续采样获取相关矩阵，压缩感知类算法[84]适用于稀疏信道场景，算法精度高但是需要针对离网点做特殊处理，基于 AI 的参数估计类算法[85]可以结合专家知识，发掘数据能力，获取强鲁棒高精度结果，但是需要考虑泛化设计。

感知业务采用不同的感知性能评价指标。感知定位通常采用分辨率、精度和无模糊范围等指标，而目标检测通常采用检测概率和虚警概率等。文献[86]提出一种相干分布式天线阵列测距波形设计，推导了测距精度的理论边界。文献[87]进一步基于高精度测距实现定位，并且证明了同等定位精度下，分布式天线阵列尺寸要求更小。文献[88]推导了集中式天线和分布式天线阵列检测概率和虚警概率的闭式解，从理论上证明了分布式天线具有更优的检测性能。用于目标感知的接收 SINR 决定了目标的检测性能和定位性能，分布式天线系统处理中心处融合得到的感知信干噪比可以表示为[89]

$$\mathrm{SINR}_{\mathrm{s}}=\mathbb{E}\left|\alpha_0\boldsymbol{w}^{\mathrm{H}}\boldsymbol{A}\left(\boldsymbol{\theta}_0\right)\boldsymbol{x}\right|^2\Bigg/\left(\mathbb{E}\left[\left|\boldsymbol{w}^{\mathrm{H}}\sum_{i=1}^{L}\alpha_i\boldsymbol{A}\left(\boldsymbol{\theta}_i\right)\boldsymbol{x}\right|^2\right]+\sigma_{\mathrm{w}}^2\right) \tag{15}$$

其中 $\alpha_0$ 和 $\alpha_i$ 分别是目标和第 $i$ 个干扰源的反射系数，由其雷达散射截面决定，$1\le i\le I$，$I$ 干扰源数量；$\boldsymbol{w}\in\mathbb{C}^{LM\times 1}$ 为接收波束成形矢量，$L$ 是总 AP 数，$M$ 为每个 AP 的接收天线数；$\boldsymbol{\theta}_0=[\theta_{0,l}]_{1\le l\le L}$ 和 $\boldsymbol{\theta}_i=[\theta_{i,l}]_{1\le l\le L}$ 分别为目标和第 $i$ 个干扰源在 $L$ 个 AP 端的到达方位角；$\boldsymbol{A}\left(\boldsymbol{\theta}_i\right)$ 是第 $i$ 个散射体在收发端的导向矩阵；$\boldsymbol{x}$ 是发射信号矢量；$\sigma_{\mathrm{w}}^2$ 为噪声能量。可见，分布式天线系统的感知信干噪比增益的来源于协作信号合并增益。图 13 展示了分布式天线系统感知信干噪比与协作的 AP 数量之间的关系。可见随着协作 AP 数量的增加，感知信干噪比也在增加，提升感知检测性能。

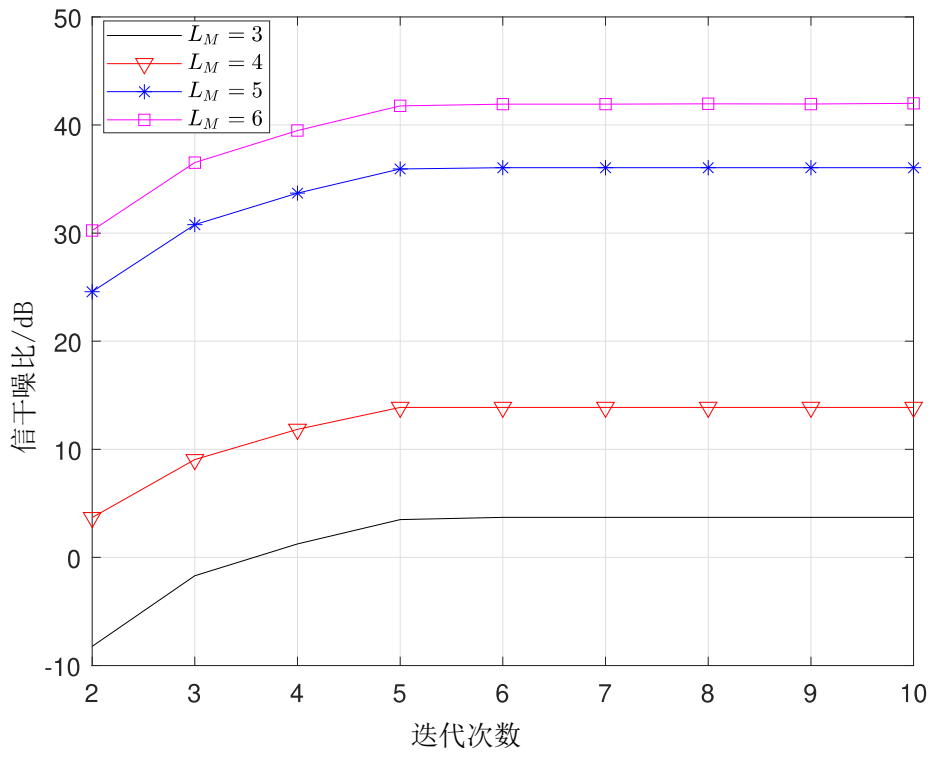

图 13 不同协作 AP 数量下的感知信干噪比收敛曲线，$L_M=3, 4, 5, 6$[89]

Fig. 13 Convergence performance for different maximum numbers of the coordinated APs, $L_M=3, 4, 5, 6$ [89]

假设区域内分布式天线节点开展协作感知，其中 $M$ 个发射机，$N$ 个接收机，第 $n$ 个接收机与第 $m$ 个发射机的基带信号模型如下

$$r_{m,n}(t)=\sqrt{\alpha_{m,n}p_m}\varsigma_{m,n}s_m\left(t-\tau_{m,n}\right)+w_{m,n}(t) \tag{16}$$

其中 $\alpha_{m,n}$ 为第 $m$ 个发射机与第 $n$ 个接收机的路径损耗；$p_m$ 为第 $m$ 个发射机的发射功率；$\varsigma_{m,n}$ 为目标的反射系数，由其雷达散射截面决定；$s_m$ 为第 $m$ 个发射机基带信号；$\tau_{m,n}$ 为第 $m$ 个发射机经感知目标到达接收机的时延；$w_{m,n}$ 为高斯白噪声，方差为 $\sigma_w^2$。待估计参数 $\boldsymbol{u}=[x,y]^{\mathrm{T}}$ 为感知目标的位置，定位估计误差可以由费雪信息矩阵的逆表示[90]

$$\boldsymbol{C}\left(\boldsymbol{u}\right)=\left\{\sum_{m=1}^{M}p_m\frac{8\pi^2B_m^2}{\sigma_w^2c^2}\sum_{n=1}^{N}\alpha_{m,n}\left|\varsigma_{m,n}\right|^2\begin{bmatrix}(x_m^{\mathrm{TX}}-x)/r_m^{\mathrm{TX}}+(x_n^{\mathrm{RX}}-x)/r_n^{\mathrm{RX}}\\(y_m^{\mathrm{TX}}-y)/r_m^{\mathrm{TX}}+(y_n^{\mathrm{RX}}-y)/r_n^{\mathrm{RX}}\end{bmatrix}\begin{bmatrix}(x_m^{\mathrm{TX}}-x)/r_m^{\mathrm{TX}}+(x_n^{\mathrm{RX}}-x)/r_n^{\mathrm{RX}}\\(y_m^{\mathrm{TX}}-y)/r_m^{\mathrm{TX}}+(y_n^{\mathrm{RX}}-y)/r_n^{\mathrm{RX}}\end{bmatrix}^{\mathrm{T}}\right\}^{-1} \tag{17}$$

其中 $B_m$ 发射信号带宽，$r_m^{\mathrm{TX}}$ 和 $r_n^{\mathrm{RX}}$ 分别表示发射机和接收机到感知目标的距离。由公式可以看出 $M$ 个发射机与 $N$ 个接收机的分布式协作感知能够为感知定位估计带来显著的费雪信息增益，包括发射机、接收机和雷达散射截面增益，进而提升感知定位精度。图 14 中展示了定位估计误差与分布式协作节点数量的关系。

可见随着协作节点数量的增加，定位估计的精度增加，估计误差减少。

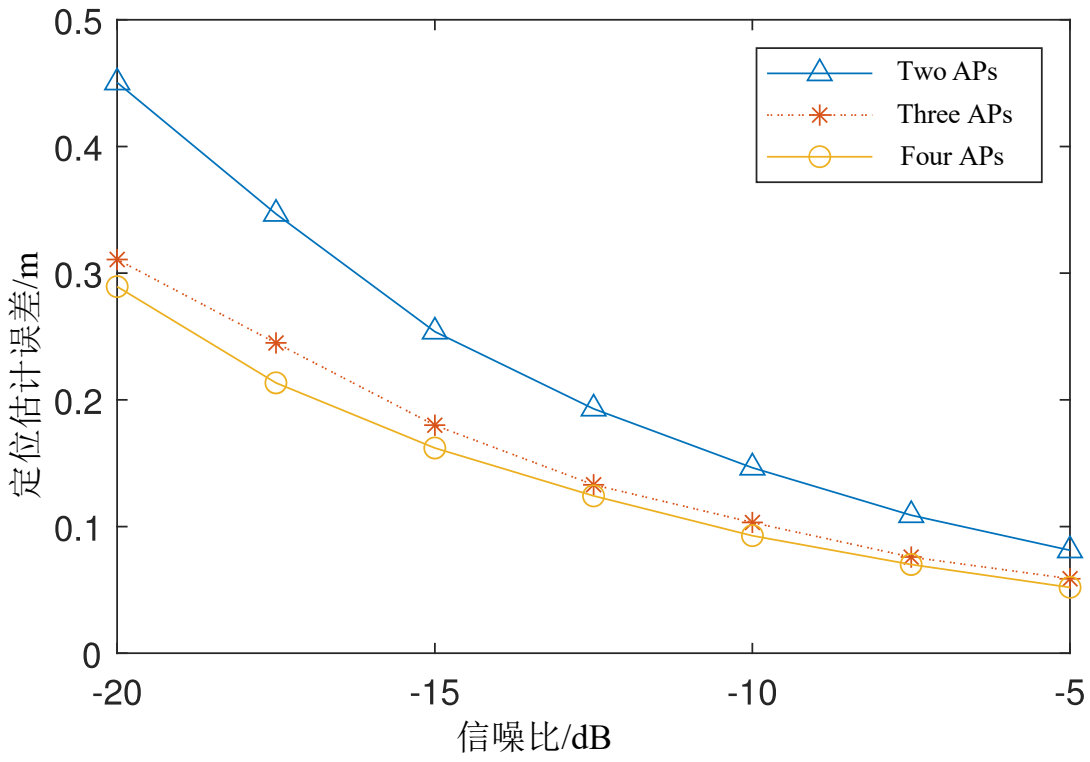

图 14 不同数量协作节点的定位估计误差随信噪比变化关系[91]
Fig. 14 The localization estimation error versus SNR for different number of collaborative nodes[91]

假设区域内分布着 $K$ 个传感器，每个传感器的检测结果传输至分布式 MIMO 系统，系统的天线阵列在空间上分布式部署，但是通过回传连接到融合中心，基于奈曼-皮尔逊准则，分布式 MIMO 系统的检测器可以设计为[88]

$$T_{\mathrm{D}}\left(\boldsymbol{Z}_{\mathrm{D}}\right)=\sum_{k=1}^{K} a_k d_{k,j_k} \Re\left(\boldsymbol{z}_{\mathrm{D},k}^{\mathrm{H}} \boldsymbol{C}_{\mathrm{D},k}^{-1} \boldsymbol{u}_k\right) \underset{\mathcal{H}_0}{\overset{\mathcal{H}_1}{\gtrless}} \tilde{\gamma} \tag{18}$$

其中 $a_k$ 是第 $k$ 个传感器的检测性能，$d_{k,j_k}$ 表示第 $k$ 个传感器和它传输的第 $j_k$ 个融合中心之间的距离因子，$\Re(\cdot)$ 表示实部运算符，$\boldsymbol{z}_{\mathrm{D},k}$ 表示第 $k$ 个传感器在分布式系统的混合波束成形器中的输出向量，$\boldsymbol{Z}_{\mathrm{D}}=[\boldsymbol{z}_{\mathrm{D},k}]_{1\le k\le K}$ 为 $K$ 个传感器在分布式系统的混合波束成形器中的输出矩阵，$\boldsymbol{C}_{\mathrm{D},k}$ 表示等效噪声协方差矩阵，$\boldsymbol{u}_k$ 表示是第 $k$ 个传感器的传输信号向量，$\tilde{\gamma}$ 表示检测阈值。分布式 MIMO 系统检测器的虚警概率 $P_{\mathrm{FA}}$ 和检测概率 $P_{\mathrm{D}}$ 分别表示为

$$P_{\mathrm{FA}}=Q\left(\frac{\tilde{\gamma}-\mu_{T_{\mathrm{D}}|\mathcal{H}_0}}{\sigma_{T_{\mathrm{D}}|\mathcal{H}_0}}\right),\ P_{\mathrm{D}}=Q\left(\frac{\tilde{\gamma}-\mu_{T_{\mathrm{D}}|\mathcal{H}_1}}{\sigma_{T_{\mathrm{D}}|\mathcal{H}_1}}\right) \tag{19}$$

其中 $Q(\cdot)$ 表示高斯 Q 函数，$\mu_{T_{\mathrm{D}}|\mathcal{H}_0}$ 和 $\sigma^2_{T_{\mathrm{D}}|\mathcal{H}_0}$ 表示 $\mathcal{H}_0$ 假设下 $T_{\mathrm{D}}$ 的均值和方差，$\mu_{T_{\mathrm{D}}|\mathcal{H}_1}$ 和 $\sigma^2_{T_{\mathrm{D}}|\mathcal{H}_1}$ 表示 $\mathcal{H}_1$ 假设下 $T_{\mathrm{D}}$ 的均值和方差。考虑虚警和漏检两种错误情况，进一步采用错误概率评估分布式 MIMO 系统检测性能

$$P_{\mathrm{e}}=\epsilon\left(1-P_{\mathrm{D}}\right)+(1-\epsilon)P_{\mathrm{FA}} \tag{20}$$

其中 $\epsilon$ 表示信号存在的先验概率。由于分布式 MIMO 系统的检测器利用了空间增益，分布式系统相比于集中式系统具备更低的错误概率，图 15 中展示了集中式 MIMO 系统和分布式 MIMO 系统理论和仿真的错误概率曲线，可以发现在-10 dB 信噪比下，分布式 MIMO 系统的检测错误概率约比集中式 MIMO 系统低 15%，分布式系统的检测性能更优。

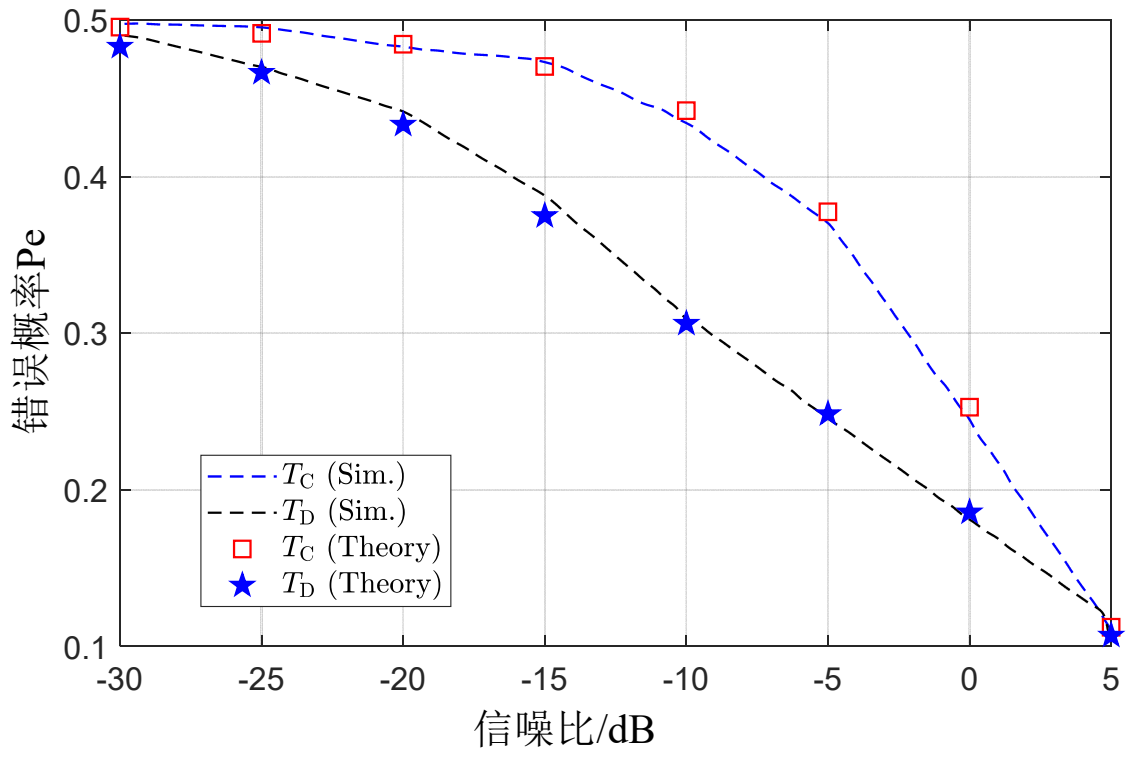

图 15 分布式 MIMO 系统和集中式 MIMO 系统的检测错误概率对比[88]
Fig. 15 Comparison of detection error probabilities between distributed and centralized MIMO systems [88]

分布式天线在通感一体化系统中更具备内生优势。分布式天线架构通过将天线分散部署于不同位置，可以提高分集能力、提升信号覆盖范围、降低所需传输功率以及增强信道多样性。基于分布式天线架构的通信感知相比于传统感知，可发挥通信系统多节点优势，利用分布式部署解决多径、遮挡等单节点感知难以解决的问题。分布式架构使得基站间具备了联合处理能力[81]，同一区域大量 AP 节点的部署使得 AP 节点与感知目标更接近，出现视距链路概率更高，降低了感知难度[86]。文献[92]总结与分析了不同的通感评估准则，包括容量-失真理论、基于均方误差的通感性能边界理论、基于感知等效速率的通感性能边界理论以及面向目标检测的通感性能边界理论，支撑了通感一体化系统优化设计。文献[93]在无蜂窝大规模 MIMO 系统下，考虑了目标位置的不确定性，推导得到了通感可达区域。文献[94]设计了分布式天线通感一体化方案，并在理论上证明了该方案的误码率、模糊函数、检测概率和虚警概率等性能优于传统的集中式通感一体化系统。进一步发挥网络节点分布式部署和泛在感知的优势，还需要解决通感网络中分布式协作感知、通感资源竞争、干扰管理控制三大关键问题，挖掘分布式天线架构的全局协作潜力。

同步问题是分布式协作感知的核心问题之一，分布式天线之间由于非共本振引起的时延、频偏误差将造成感知性能下降。基于校准的同步方案可以估计网络内分布式节点对的时延、频偏误差[95-96]，进而实现全局同步，但是该方案需要较长时间的初始化校准，时间开销较大。基于交叉天线互相关的信号处理方案可以抵消天线对之间的时延、频偏误差，但是会产生镜像分量，需要对感知算法做相应的修正[97-98]。信号联合处理是分布式协作感知的另一核心问题，可以采用先解算时延、多普勒等感知参数再联合处理的方案，该方案降低了前传开销，但是需要解决多视角和参数匹配问题[99-100]。也可以采用先联合处理再解算感知参数的方案，该方案可以获取相参增益，提升探测距离和感知性能。文献[101]提出了一种利用多普勒实现相干反向投影的算法，达到厘米级定位精度。

由于分布式天线架构的硬件共享，通信和感知在时空频域和功率域存在资源竞争。文献[102]提出了一种基于拉格朗日方法的功率分配方案最大化感知信干噪比和通信和速率，文献[103]联合优化下行通感双功能预编码、上行接收预编码和上行用户的发射功率，在保障通感性能的基础上实现分布式系统的功耗最小化与通信和速率最大化。文献[104]提出一种集成主观用户需求和客观环境需求的 6G 通感算资源按需调配架构，高效解决资源竞争和调度问题。此外，分布式架构下通感资源竞争解决的前提是初始关联，用户和 AP 节点的初始关联以及感兴趣目标和 AP 节点的初始关联也将成为一个重要的研究课题[105]。

根据通信和感知的业务需求不同，基于分布式天线架构的通感系统将存在上行用户干扰、下行用户干扰、感知信号干扰三种干扰类型，干扰的管理和控制将成为制约通感性能的关键问题。时分和频分多址是干扰管控的基本策略[106]。空分多址是干扰管控的常见手段，但是在干扰较强时，通感系统性能将降低。文献[107]给出了以感知为中心和以通信为中心的接收波束成形的基本结论，并且给出了一种基于凸优化的通感一体联合优化方案。速率分拆多址正成为一个统一的和强大的多天线干扰管理新策略，其核心思想是在发射器上将消息分割为公共和私有流，在接收器上连续干扰取消。文献[108]设计了通信流和雷达序列的消息分割和预编码器，以使每个天线功率约束下的加权和率和雷达波束模式近似均方误差最小化，但是感知的设计目标仍局限于感知信号检测。文献[109]则更进一步考虑感知目标定位误差界，采用连续凸逼近逼近最优帕累托边界，结果表明速率分拆多址可以实现多天线系统干扰管控，提升通感系统性能。

表 5 总结了分布式天线在无线通信及无线感知领域的主要研究内容及代表性论文。

**表 5 分布式天线系统代表性论文**
**Tab. 5 Representative works of cell-free antenna system**

| 研究内容 | | 代表性论文 |
|---|---|---|
| 通感一体 | 通感性能 | 性能分析[82-85]，性能分析[86-93] |
| | 协同感知 | 同步问题[94-98]，联合处理[99-101] |
| | 资源竞争 | 资源调度[102-104]，初始关联[81][105] |
| | 干扰管控 | 时分和频分多址[106]，空分多址[107]，速率分拆多址[108-109] |

### 4.4 分布式天线通信与感知小结

分布式天线系统通过在区域内部署多个天线单元，能够在广泛区域内提供均匀的无线服务覆盖，实现了信号覆盖的连续性和质量的提升，尤其适用于学校、高铁站和商超等热点区域。分布式天线系统协作感知可利用目标散射特性，获得更强回波能量，可以内生实现定位和检测等，可以助力万物智联、数字孪生等 6G 愿景。但是分布式天线系统部署成本和实现复杂度较高，需要解决系统同步、下行相干传输、前传信息交互等技术难题。为实现通感一体化，未来还需要设计高效的协同感知方案，解决资源竞争和干扰管控问题。

## 5 可移动天线通信与感知

### 5.1 信道建模及信道估计

#### 5.1.1 信道建模

对于传统的固定式天线，信道模型只需要刻画给定位置的收发天线之间的信道响应。相比之下，可移动天线的空间位置可以灵活改变，因此信道模型需要刻画天线在收发区域内不同位置的信道响应。为此，文献[38]提出了一种基于场响应的信道模型，可以表征无线信道在收发区域的连续变化。具体而言，在远场条件下，记发射区域和接收区域的信道多径数量分别为 $L_\mathrm{t}$ 和 $L_\mathrm{r}$。相应地，第 $j$ 条发射路径的俯仰出发角和方位出发角分别记为 $\theta_{\mathrm{t},j}$ 和 $\phi_{\mathrm{t},j}$，$1\le j\le L_\mathrm{t}$；第 $i$ 条接收路径的俯仰到达角和方位到达角分别记为 $\theta_{\mathrm{r},i}$ 和 $\phi_{\mathrm{r},i}$，$1\le i\le L_\mathrm{r}$。由此，可以得到发射多径的波矢量 $\boldsymbol{k}_{\mathrm{t},j}=[\cos\theta_{\mathrm{t},j}\cos\phi_{\mathrm{t},j},\cos\theta_{\mathrm{t},j}\sin\phi_{\mathrm{t},j},\sin\theta_{\mathrm{t},j}]^\mathrm{T}$ 和接收多径的波矢量 $\boldsymbol{k}_{\mathrm{r},i}=[\cos\theta_{\mathrm{r},i}\cos\phi_{\mathrm{r},i},\cos\theta_{\mathrm{r},i}\sin\phi_{\mathrm{r},i},\sin\theta_{\mathrm{r},i}]^\mathrm{T}$。记发射天线位置在三维局部直角坐标系内的坐标为 $\boldsymbol{t}=[x_\mathrm{t},y_\mathrm{t},z_\mathrm{t}]^\mathrm{T}$，接收天线位置在三维局部直角坐标系内的坐标为 $\boldsymbol{r}=[x_\mathrm{r},y_\mathrm{r},z_\mathrm{r}]^\mathrm{T}$。在发射端和接收端分别定义信道场响应矢量如下

$$\begin{aligned}\boldsymbol{g}(\boldsymbol{t})&=\left[\exp\left(\mathrm{j}2\pi/\lambda\cdot\boldsymbol{k}_{\mathrm{t},1}^\mathrm{T}\boldsymbol{t}\right),\exp\left(\mathrm{j}2\pi/\lambda\cdot\boldsymbol{k}_{\mathrm{t},2}^\mathrm{T}\boldsymbol{t}\right),\cdots,\exp\left(\mathrm{j}2\pi/\lambda\cdot\boldsymbol{k}_{\mathrm{t},L_\mathrm{t}}^\mathrm{T}\boldsymbol{t}\right)\right]\\ \boldsymbol{f}(\boldsymbol{r})&=\left[\exp\left(\mathrm{j}2\pi/\lambda\cdot\boldsymbol{k}_{\mathrm{r},1}^\mathrm{T}\boldsymbol{r}\right),\exp\left(\mathrm{j}2\pi/\lambda\cdot\boldsymbol{k}_{\mathrm{r},2}^\mathrm{T}\boldsymbol{r}\right),\cdots,\exp\left(\mathrm{j}2\pi/\lambda\cdot\boldsymbol{k}_{\mathrm{r},L_\mathrm{r}}^\mathrm{T}\boldsymbol{r}\right)\right]\end{aligned} \tag{21}$$

其中发射场响应矢量的每一个元素表示相应的发射路径在天线位置 $\boldsymbol{t}$ 的信道响应相对于参考点（即发射区域坐标原点）的相位差，接收场响应矢量的每一个元素表示相应的接收路径在天线位置 $\boldsymbol{r}$ 的信道响应相对于参考点（即接收区域坐标原点）的相位差。进一步定义 $L_\mathrm{r}\times L_\mathrm{t}$ 维路径响应矩阵 $\boldsymbol{\Sigma}$，其中第 $i$ 行、第 $j$ 列的元素表示从发射区域参考点到接收区域参考点的第 $j$ 条发射路径与第 $i$ 条接收路径的之间信道响应复系数。由此可以得到发射天线位置 $\boldsymbol{t}$ 到接收天线位置 $\boldsymbol{r}$ 之间的信道系数为

$$h(\boldsymbol{t},\boldsymbol{r})=\boldsymbol{f}(\boldsymbol{r})\boldsymbol{\Sigma}\boldsymbol{g}(\boldsymbol{t}) \tag{22}$$

该信道模型可以拓展到多天线系统，记 $N$ 个发射天线位置坐标集合为 $\boldsymbol{T}=[\boldsymbol{t}_1;\boldsymbol{t}_2;\cdots;\boldsymbol{t}_N]$，$M$ 个接收天线位置集合为 $\boldsymbol{R}=[\boldsymbol{r}_1;\boldsymbol{r}_2;\cdots;\boldsymbol{r}_M]$。则发射端信道场响应矩阵定义为 $\boldsymbol{G}(\boldsymbol{T})=\left[\boldsymbol{g}(\boldsymbol{t}_1),\boldsymbol{g}(\boldsymbol{t}_2),\cdots,\boldsymbol{g}(\boldsymbol{t}_N)\right]$，接收端信道场响应矩阵定义为 $\boldsymbol{F}(\boldsymbol{R})=\left[\boldsymbol{f}(\boldsymbol{r}_1),\boldsymbol{f}(\boldsymbol{r}_2),\cdots,\boldsymbol{f}(\boldsymbol{r}_M)\right]$。则收发天线之间的信道响应矩阵可以表示为

$$\boldsymbol{H}(\boldsymbol{T},\boldsymbol{R})=\boldsymbol{F}(\boldsymbol{T})\boldsymbol{\Sigma}\boldsymbol{G}(\boldsymbol{R}) \tag{23}$$

公式(22)和(23)所示的基于场响应的信道模型将信道系数表示为收发天线位置的函数，精确刻画了信道响应在三维连续空间内的变化过程，且该信道模型兼容了传统的固定天线位置信道模型，在特定条件下可以退化为经典的视距信道模型、几何信道模型、瑞利衰落信道模型以及莱斯衰落信道模型。值得指出的是，对于天线移动范围较小（通常为几个波长量级）的无线通信系统，收发机天线之间满足远场条件，此时可以用均匀平面波建模收发机之间的多径响应，得到式(22)和(23)所示的场响应信道模型。如果天线移动区域尺寸较大或收发机间距较小，平面波模型将不再适用。例如，文献[110]考虑了视距条件下的可移动天线 MIMO 系统，采用了球面波模型对视距信道进行了近似表征，但没有考虑天线位置变化对路径响应系数的幅度影响，也没有对非视距路径进行建模。因此，在近场条件下的场响应信道建模还有待进一步研究。

除了基于场响应的信道模型，可移动天线/流体天线系统还有基于空间相关性的统计信道模型。以文献[40]中的一维可移动空间为例，接收区域是长度为 $W\lambda$ 的线段，将其均匀分成 $M$ 个天线端口，即天线备选位置。第 $m$ 个天线端口的信道响应系数服从瑞利衰落，记为复高斯信道增益 $g_m$，$1\le m\le M$，其中任意两

个天线端口的信道相关性服从 Jake 模型。对于第 $m$ 个和第 $n$ 个天线端口，二者的信道相关系数为

$$\mathrm{Corr}_{m,n}=\frac{\sigma^2}{2}J_0\left(\frac{2\pi(m-n)}{M-1}W\right) \tag{24}$$

其中 $\sigma^2$ 表示复高斯信道增益的方差，$J_0$ 是第一类零阶贝塞尔函数。为了满足上述信道相关性，文献[40]将 $M$ 个端口的信道系数参数化为

$$\begin{cases} g_1=\sigma x_0+\mathrm{j}\sigma y_0 \\ g_m=\sigma\left(\sqrt{1-\mu_m^2}x_m+\mu_m x_0\right)+\mathrm{j}\sigma\left(\sqrt{1-\mu_m^2}y_m+\mu_m y_0\right),\ 2\le m\le M \end{cases} \tag{25}$$

其中 $x_0,x_1,\cdots x_M$，$y_0,y_1,\cdots y_M$ 为独立同分布复高斯随机变量，其均值为 0，方差为 0.5。参数 $\mu_m$ 作为待定系数用来刻画信道相关性，$2\le m\le M$，设置为 $\mu_m=J_0\left(2\pi(m-1)/(N-1)\cdot W\right)$。

上述基于空间相关性的信道模型优点在于简化了复杂的信道参数，提取了不同空间位置的统计特征，以便于对可移动天线/流体天线系统进行性能分析。值得指出的是，公式(24)所示的信道相关性是场响应信道模型的一种特例，条件为在二维平面内有无穷条接收路径且路径响应系数服从独立同分布，即均匀散射条件。在三维空间内的均匀散射条件下，信道相关系数则可以表示为天线间距的 sinc 函数形式。此外，由于 $\mu_m$ 所设置的参数在非参考端口的信道相关系数不严格满足公式(24)，已有文献对上述空间相关性信道进行了改进，并且拓展到二维平面的信道相关性以及收发区域均采用可移动天线/流体天线的系统[110-113]。然而，这些基于空间相关性的统计信道模型都无法从物理环境出发建模信道相关性，并且无法在数学上论证信道相关系数的准确性。例如，公式(24)中的信道相关系数所服从的 Jake 模型，是基于均匀散射条件得到的，但实际通信系统并不一定满足这一条件。在不同的信号传播环境下，信道相关性可能是 Jake 模型以外的任意形式的函数。因此，可移动天线系统更精准的统计信道建模还有待进一步研究。

### 5.1.2 信道估计

传统的固定式天线只需要估计发射天线的固定位置与接收天线的固定位置之间的信道响应，可以直接通过测量导频信号得到。相比之下，由于可移动天线的空间位置可以灵活改变，相应的信道响应也随之发生改变，因此针对可移动天线/流体天线系统的信道估计则需要估计出发射区域每一点与接收区域每一点之间的信道响应，其中包含无穷数量的信道响应。如果采用遍历发射区域与接收区域的每一对点直接对信道响应进行测量则会带来大量的导频与时间开销，对于可移动天线系统无法实现。为此，基于上述的场响应信道模型与空间相关性信道模型，已有文献分别提出了角度域与空间域的信道测量方法。

角度域的信道估计方法基于角度域的等效多径稀疏性，通过估计多径分量信息的方法重构发射区域与接收区域之间的完整信道状态信息（channel state information, CSI），可有效降低信道估计的导频开销。具体而言，相比于空间域中无穷数量的信道响应，多径分量的数量是有限的，表明角度域的路径响应相比于空间域的信道响应具有稀疏性；进一步，有限的发射/接收区域将导致受限的角度分辨率，即具有相似发射角/到达角的发射/接收路径可以有效地等效为一条主要路径，进一步增强了角度域的路径响应的稀疏性。基于以上两点原因，可移动天线系统具有角度域等效多径稀疏性。由此，可以利用稀疏信号恢复的方法，通过有限次的信道测量对多径分量信息进行估计，即每一个多径分量的发射角、到达角以及信道复系数。并进一步利用场响应信道模型，可重构出发射区域任意一点与接收区域任意一点之间的信道响应。基于以上方法，文献[114]首次提出了连续发射接收压缩感知方法对可移动天线系统进行信道估计。其首先令发射/接收端天线移动至各自区域内选定的位置点进行有限次的信道测量，基于信道测量结果利用压缩感知的方法即可恢复出所有多径分量的发射角与到达角；进一步，基于估计的全部发射角与到达角，利用最小二乘估计即可恢复出所有多径分量的信道复系数；最后，利用场响应信道模型，可重构出收发区域任意点对之间的信道响应。文献[114]通过有限次的信道测量以较高的准确性恢复出了发射区域与接收区域之间的完整 CSI，然而其对多径分量信息的串行估计会带来较大的累计误差，因此针对可移动天线的信道估计准确性可以进一步提高。为此，文献[115]同样基于压缩感知，提出了一种对多径分量信息联合恢复的信道估计方法。具体而言，其每一次迭代可以直接恢复出多径分量的发射角、到达角以及信道复系数。最后，同样根据恢复出的多径分量信息，利用场响应模型重构出发射区域与接收区域之间的完整 CSI。文献[115]所提的多径分量信息联合估计方法可消除串行估计带来的累积误差，进一步提升信道估计的准确性。

除此之外，空间域的信道估计方法则是利用可移动天线/流体天线的区域中较强的信道空间相关性，通

过在空间中相距较远、空间相关性较低的位置进行信道测量，再利用回归或插值等方法即可重构出发射区域与接收区域之间的完整 CSI，同样达到降低导频开销的目的。空间域信道估计方法首次在文献[116]中提出，称为跳跃使能的线性最小均方误差信道估计方法。具体而言，其首先令发射/接收端可移动天线/流体天线移动到等距的位置点进行信道测量，并通过线性最小均方误差估计出收发端之间的信道响应；进一步，基于空间域具有强相关性的假设[40]，即在较短距离内信道响应是慢变的，令未被测量区域的信道响应与最近的测量位置点的信道响应相同，即可重构出发射区域与接收区域之间完整的 CSI。此信道估计方法的准确性依赖于信道在空间域具有强相关性的假设。具体而言，一方面，相邻测量位置点相距较远将导致空间相关性降低，从而降低信道估计的准确性；另一方面，相邻测量位置点相距较近则会在提升信道估计准确性的同时带来过高的导频开销。文献[117]提出的连续贝叶斯重构方法则是将可移动天线/流体天线的信道建模为随机过程，进而基于信道空间强相关性的假设，根据可移动天线/流体天线在不同位置点有限的信道测量利用贝叶斯回归估计出完整的 CSI。具体而言，首先在离线训练阶段，将可移动天线/流体天线的信道建模为高斯随机过程，并根据最大后验方差准则选出信道测量位置并确定权重向量；进一步在线回归阶段则可以直接根据在预定位置的信道测量结果通过权重向量重构完整的 CSI。值得注意的是，文献[116-117]中仅考虑了接收端采用一维可移动天线/流体天线的场景，将其拓展至发射端与接收端均采用可移动天线/流体天线或天线二维移动的场景则可能会给离线训练阶段带来较高的计算复杂度。

文献[114-115]所提的角度域信道估计方法利用了信道路径的角度域稀疏性，可以在较小的信道测量开销下恢复完整的 CSI，但该方法依赖于信道模型的准确性。在实际系统中，由于天线单元耦合、传播环境等复杂因素，可能会导致平面波假设不成立，因此信道估计误差可能较大。相比之下，文献[116-117]所提的空间域信道估计方法不依赖于特定的信道模型，因此适用性更广以及误差容忍度更高。但是，这类信道估计方法依赖于信道的空间相关性假设，在实际系统中很难获得准确的相关性关系，而且需要更高的信道测量开销才能达到较好的信道估计性能。在未来研究中，可以进一步考虑两种方法的结合，以实现信道估计精度和信道测量开销的最优折中。此外，基于数据驱动的信道估计方法具有更广的适应性，根据大规模信道测量数据预训练学习模型，形成部分信道测量数据到完整 CSI 的直接映射。上述信道估计方法的模型建立、协议和算法设计以及在实际系统中的实现值得进一步研究。

## 5.2 性能分析

### 5.2.1 空间分集增益

由于无线信道中各多径分量的叠加，收发端之间的信道增益会随着发射/接收端天线位置的改变而发生变化。对于传统的固定式天线，多径随机叠加会导致收发端之间的信道处于深衰落状态，从而严重影响通信质量。相比之下，可移动天线/流体天线系统则可通过灵活配置天线位置，充分利用信道空间变化获得空间分集增益[37]。具体而言，当获得了发射区域与接收区域间完整的 CSI 后，可移动天线/流体天线系统可以使天线移动至信道增益较高的位置，从而提升接收信号功率及接收信噪比；除此之外，天线可以移动至干扰发射机信道增益最小的位置，从而提升接收信干噪比。由此，即使是对于单个可移动天线/流体天线的无线通信系统，同样可以实现有效信号增强以及干扰抑制。图 16 为接收端多径数量为 6 的情况下信道增益随着接收天线位置变化图像。可以发现，在波长量级的范围内，信道增益的变化范围可以达到 40 dB 以上，这证明了天线的局部移动对提升空间分集增益的巨大潜力。

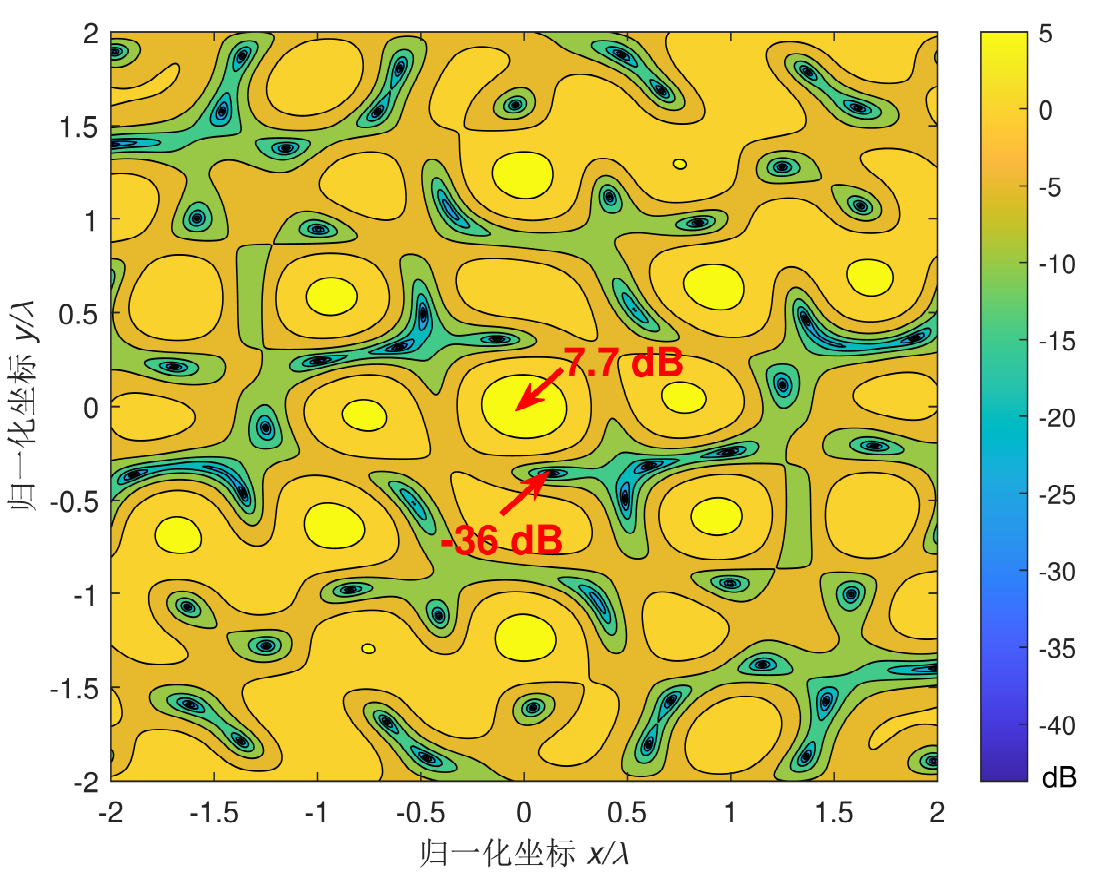

图 16　接收端信道增益随天线位置变化图
Fig. 16　The channel gain pattern at the receiver under different receive antenna positions

目前已有文献证明了可移动天线/流体天线系统带来的空间分集增益。文献[38]对二维平面内运动的可移动天线的单输入多输出（single-input single-output, SISO）系统所能获得的最大信道增益及接收信噪比进行了分析，推导得出最大信道增益的累积分布函数的近似闭式表达式，并进一步计算信道中不同多径数量下的中断率。其分析结果表明，可移动天线系统的空间分集增益来自于多径叠加，当多径数量增加时，可移动天线系统能获得的最大接收信噪比显著提升。在多径数量足够多时，所考虑的可移动天线系统甚至可以获得与单输入多输出（single-input multiple-output, SIMO）数字波束赋形系统相近的接收信噪比。文献[40][118]基于瑞利信道下的空间相关性模型，分析了一维线性运动的可移动天线/流体天线的 SISO 系统性能。其中，文献[40]推导了系统中断率及其上界，文献[118]则进一步推导了系统中断率的近似闭式表达式及所能获得的分集增益。文献[40] [118]的推导结果均证明即使采用单个可移动天线/流体天线的系统，其中断率也可以达到与采用 MRC 接收的固定式天线阵列相近的性能。除接收信号功率提升外，通过优化可移动天线位置也可以达到干扰抑制的目的。文献[37]分析了接收端采用可移动天线的 SISO 场景的干扰抑制性能，即存在额外的单个干扰源条件下所能达到的最大信干噪比。其结果显示通过优化天线位置，可以使信干噪比接近于无干扰源条件下的信噪比，说明可移动天线系统可以在期望信号功率几乎无损失的条件下消除干扰。除此之外，可移动天线也可与智能超表面（reconfigurable intelligent surface, RIS/intelligent reflecting surface, IRS）相结合，为通信系统带来额外的空间分集增益。具体而言，文献[119]考虑了发射天线，接收天线与 RIS/IRS 可以旋转的场景，推导了莱斯信道下系统遍历容量的上界，并在此基础上求解出了发射/接收天线与 RIS/IRS 的最佳旋转角度。其结果指出通过优化 RIS/IRS 的旋转角度将显著提升通信系统的性能。除旋转外，文献[120]对由线性可移动阵列组成的 RIS/IRS 进行了分析，并指出通过优化 RIS/IRS 各阵元的位置，可以有效消除由 RIS/IRS 端不同发射角与到达角带来的相位偏移的影响，显著提升系统的可达率。

### 5.2.2　空间复用增益

在具有空间复用能力的 MIMO 系统中，可移动天线可以为其提供更高的空间复用增益，这是由于通过改变可移动天线的位置，可以重塑信道矩阵，有效改善信道状态[37]。具体而言，通过优化可移动天线的位置，可以重塑信道矩阵并改变其奇异值，进而提升 MIMO 系统的信道容量。例如，在低信噪比条件下，单流传输信道容量最大，因此可以通过优化可移动天线位置增大信道矩阵的最大奇异值以提升信道容量；在高信噪比条件下，可以基于注水法调整信道矩阵的各个奇异值，从而提升信道容量。图 17 对比了 6 发 6 收 MIMO 系统下可移动天线与固定位置天线的信道容量。可以看到，相较于固定位置天线系统，可移动天线系统的 MIMO 信道容量显著提升，尤其在高信噪比以及多径数量较多的情况下性能增益更高。

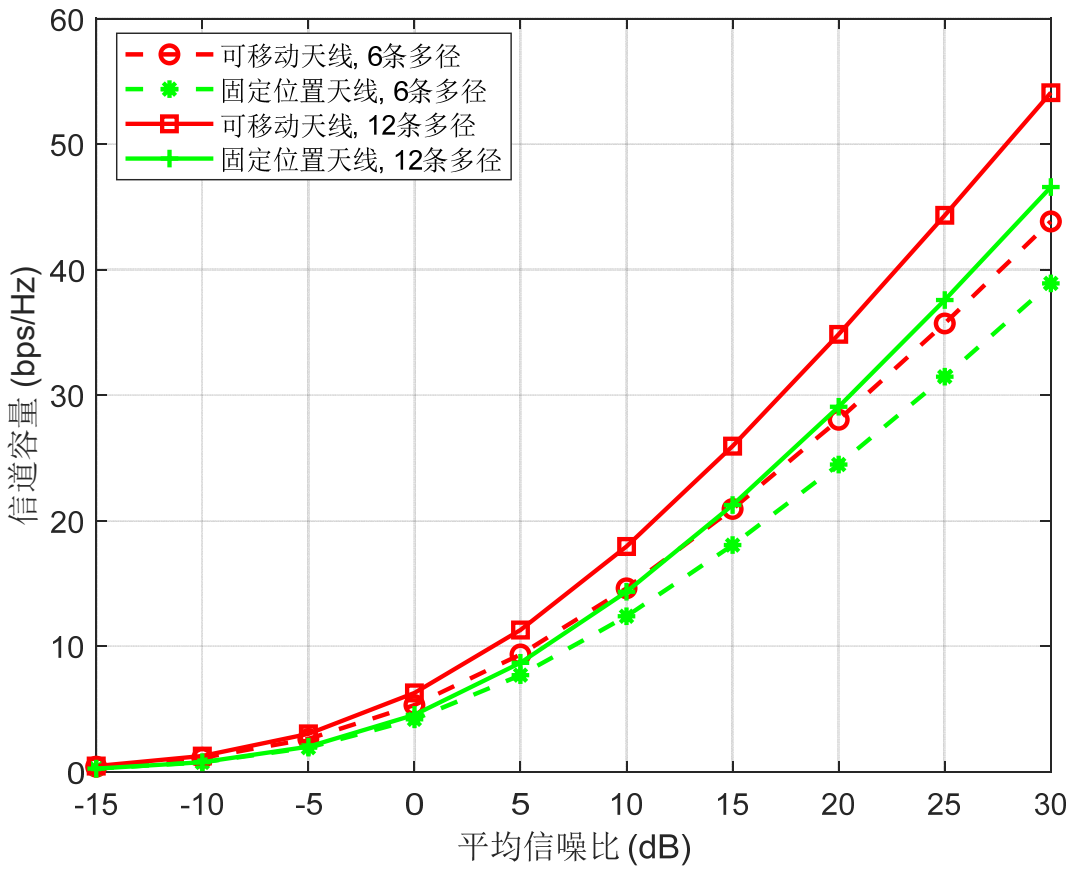

图 17　可移动天线和固定位置天线 MIMO 信道容量对比
Fig. 17　The comparison of channel capacity of movable antenna and fixed-position antenna systems

目前已有文献证明了可移动天线带来的空间复用增益。在提升信道容量方面，文献[121]考虑可移动天线 MIMO 系统下的信道容量优化问题。具体而言，其考虑对发射/接收端天线位置以及发射信号的协方差

矩阵联合优化，实现系统信道容量的最大化。文献[121]证实相比于传统的固定式天线系统，可移动天线系统可以通过优化天线位置重塑信道矩阵，从而显著提高系统的信道容量。进一步，文献[122]基于统计 CSI，对 MIMO 系统中发射/接收端可移动天线位置及发射信号协方差矩阵联合优化，实现了可达率最大化。文献[123]针对多用户上行传输场景，其中每个用户采用单个传统固定式天线以及基站采用多个可移动天线，在每个用户最大发射功率的约束下，通过对可移动天线的位置、用户发射功率以及接收端组合矩阵的联合优化，实现了每个用户最小可达率的最大化。文献[124-125]均考虑多用户下行传输场景，其中天线配置与文献[123]中相同。文献[124]在基站最大发射功率的约束下，通过对发射天线的位置以及发射端波束赋形矢量的联合优化，实现了用户总可达率的最大化。不同于文献[124]，文献[125]假设基站与用户间视距链路并采用 RIS/IRS 辅助通信，通过联合优化发射天线位置，波束赋形矢量以及 RIS/IRS 的反射系数，在基站最大发射功率的约束下实现了用户总可达率的最大化。文献[126]对基站采用单个固定位置天线，用户采用单个可移动天线/流体天线的下行传输场景进行了优化，其通过优化用户端可移动天线/流体天线位置及各用户的功率分配，分别在正交多址接入与非正交多址接入系统中实现了用户和速率的最大化。除此之外，可移动天线/流体天线系统可以通过灵活配置天线位置而抑制多用户干扰，文献[127]提出了一种新的多址接入方式，即流体天线多址接入。具体而言，相比于传统的需要进行复杂信号处理与编码的多址接入方式，流体天线多址接入仅需要将各个用户的天线移动到信干比区域即可有效抑制多用户干扰，而无需复杂的信号处理。文献[127]推导了流体天线多址接入系统的中断率上界的闭式表达式以及平均中断率下界。其分析结果表明通过在每个用户部署可移动天线/流体天线，可以有效降低系统的中断率，提升系统容量。

除信道容量外，通过优化可移动天线位置并重构信道矩阵，可以显著提高信道矩阵增益以降低所需发射功率及提升接收信噪比。具体而言，文献[128-130]考虑基站配备固定式天线阵列，每个用户采用单个可移动天线的场景。其中文献[128]针对多用户的上行传输场景，在每个用户满足最小可达率的约束下，通过对可移动天线的位置、每个用户的发射功率以及接收端组合矩阵的联合优化，实现了用户总发射功率的最小化。文献[129]考虑了多用户的下行传输场景中发射功率优化的问题。在每个用户最小信干噪比的约束下，通过联合优化每个可移动天线位置以及发射端波束赋形矢量，可以实现总发射功率的最小化。文献[130]针对下行传输场景，假设基站每个天线服务单个用户，通过优化每个用户的功率分配及可移动天线/流体天线位置，实现了平均能量效率的最大化。文献[131-132]考虑基站配置多个可移动天线以及用户配置单个固定位置天线的场景，其中文献[131]针对采用迫零接收的上行传输，通过优化接收端可移动天线的位置，在每个用户最小速率的约束下实现了总发射功率的最小化。文献[132]针对下行传输场景，对发射端可移动天线位置以及波束赋形矢量进行联合优化，在每个用户最小信干噪比的约束下实现了总发射功率的最小化。值得说明的是，文献[132]还考虑了可移动天线的实际硬件限制，将可移动天线的移动建模为离散运动，即在每个可移动天线只能位于发射区域的离散网格点上的约束下对其位置进行优化。文献[133]分析了可移动天线对多点协同接收带来的性能增益。具体而言，发射端配备线性可移动天线阵列，为多个接收端发送共同信息，多个接收端再利用最大比合并对共同信息解码。通过联合优化发射端可移动天线位置及波束赋形矢量，在最大发射功率的约束下实现了接收信噪比的最大化。

在安全通信领域，通过优化发射天线的位置，使窃听者的信道处于深衰落状态，提升物理层安全[134-135]。文献[134]针对发射端采用多个可移动天线，用户与窃听者采用单个固定天线的场景，通过联合优化发射端可移动天线位置以及波束赋形矢量，分别实现了发射功率最小化以及保密速率最大化。进一步，文献[135]考虑多个可移动天线的发射端与单个固定位置天线的用户以及多个窃听者的场景，其中每个窃听者采用单个固定位置天线，通过联合优化发射端可移动天线位置以及波束赋形矢量，在最大发射功率的约束下实现了保密速率最大化。

#### 5.2.3 灵活波束赋形

传统的固定式天线阵列几何结构是固定的，仅通过调整天线波束赋形矢量形成的波束不够灵活，使波束赋形的性能受限。与之相比，可移动天线/流体天线阵列除了优化波束赋形矢量外，还可以优化各个天线阵元的位置，从而实现对天线阵列的几何结构以及波束赋形矢量的联合设计，进行灵活波束赋形[37]。具体而言，可移动天线/流体天线阵列可以通过优化各天线阵元的位置改变导向矢量。一方面，如图 18 (a)所示的干扰调零，通过优化天线阵元的位置可以实现期望方向的导向矢量与干扰方向的导向矢量几乎正交，从而在期望方向上获得较大的波束赋形增益的同时在干扰方向上几乎无增益；另一方面，如图 18 (b)所示的多波束赋形，优化天线阵元位置可以增强不同方向的导向矢量之间的相关性，从而可以在不同的方向上均

获得较大的波束赋形增益。

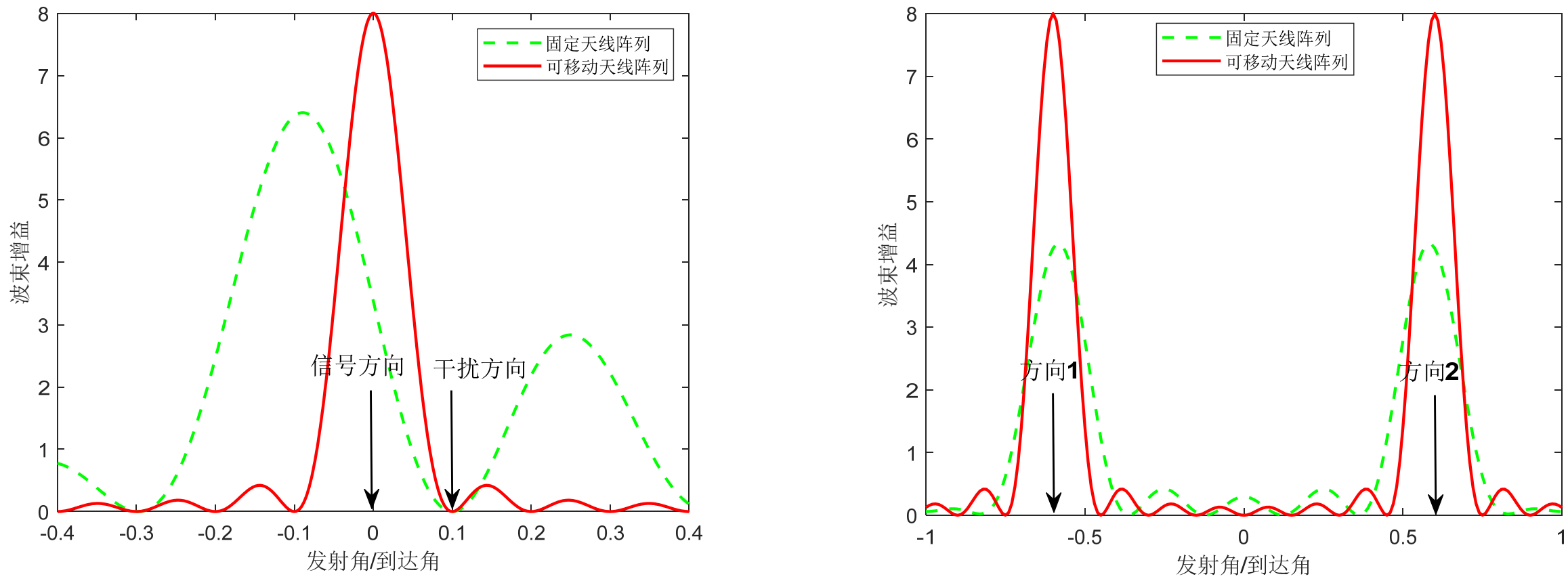

(a) 干扰调零 (b) 多波束赋形

图 18 可移动天线阵列和固定天线阵列波束对比图

Fig. 18 Comparison of beam patterns between the movable antenna array and fix-position antenna array

文献[136-137]证实了可移动天线阵列在波束赋形方面的巨大潜力。首先在文献[136]中考虑了各阵元沿着半圆形路径移动的天线阵列，并通过差分进化算法以优化各个可移动天线阵元的位置，以产生多种天线方向图，证明了通过改变各天线阵元的位置可以实现更加灵活的波束赋形。文献[138]考虑了线性可移动天线阵列的干扰抑制问题。具体而言，在给定阵元数量以及干扰方向的条件下，通过联合设计天线位置矢量（由各阵元的位置决定）以及波束赋形矢量，使期望方向的导向矢量与所有干扰方向的导向矢量正交，在期望方向获得全部波束赋形增益的条件下实现了所有干扰方向零波束赋形增益。同时，文献[138]得到了天线位置矢量与波束赋形矢量的闭式最优解，且波束赋形矢量的最优解满足恒模约束，表明可移动天线阵列仅需要模拟波束赋形即可实现对干扰的完全抑制。因此，文献[138]证明可移动天线阵列相较于传统的固定式天线阵列在干扰抑制方面具有巨大优势。除此之外，文献[137]考虑了线性可移动天线阵列的多波束赋形问题。具体而言，通过联合优化天线位置矢量与波束赋形矢量，考虑了在干扰方向上最大功率的限制下，多个期望方向上的最小波束赋形增益最大化的问题。并提出天线位置矢量与波束赋形矢量的交替优化算法，得到多波束赋形问题的次优解。其结果证明可移动天线阵列相较于传统的固定式天线阵列在多个期望方向上的波束赋形增益具有显著提高。

### 5.2.4 无线感知及通感一体化

除无线通信领域外，可移动天线在无线感知领域同样具有巨大潜力，其空间位置改变等效增大了天线孔径，可以有效提升感知精度。因此，可移动天线在雷达、定位、导航等系统得到了初步应用。具体而言，在目前的单天线惯性导航/全球卫星导航系统中，目标较低的移动性会导致其可观测性下降，使导航系统难以提供可靠的定位。尽管采用多天线的导航系统可以解决此问题，但天线数量的增加会导致成本上升。为此，通过将单个可移动天线与导航系统相结合，令可移动天线在定位目标上以特定形式移动可以显著提高目标的可观测性，在保证低成本的条件下提升导航系统定位精度。且文献[139]指出，可移动天线仅需 0.15 m/s 的移动速度即可显著提升导航系统性能。对于室内定位，通过接收端可移动天线的运动产生多普勒频移，并采用单个伪卫星配合多普勒频移值即可实现室内定位。文献[140]指出此多普勒定位方法具有达到室内分米级定位的潜力。在方向估计领域，可移动天线的移动特性使单个天线的方向估计成为了可能[140]。具体而言，传统的方向估计需要以多个空间上分离的固定式天线为基础，通过测量各天线间的相位差及幅度差以实现方向估计。而单个可移动天线可以通过移动至不同位置以获取相位差并进行方向估计。与传统的多天线方向估计方法相比，单个可移动天线方向估计具有成本低、无天线耦合效应等优势。在雷达成像领域，文献[141]采用动态超表面天线作为发射端，单个可移动天线作为接收端，配合范围迁移算法形成成像雷达，弥补了动态超表面天线无法产生均匀辐射图的缺陷，并实现了三维图像重构。文献[142]采用设计了发射端与接收端均采用单个可移动天线的成像雷达，其利用可移动天线可以移动的特性，用以灵活调整对目标图像的采样点，在较低硬件复杂度的条件下实现了对采样点的评估。

此外，文献[143]研究了可移动天线阵列的几何结构对方向角估计性能的影响。研究表明，在给定信噪

比的条件下，方向角估计的均方误差的克拉美罗界主要由多天线位置的方差决定。因此，在不改变天线数量的条件下，通过增大天线的移动范围可以显著提升角度估计性能。文献[143]进一步推导了一维可移动天线阵列的闭式最优天线位置，从而最小化角度估计误差的克拉美罗界；对于二维可移动天线阵列，给出了角度估计误差的克拉美罗界的取值范围，并通过交替优化获得了任意给定二维区域内的次优天线位置，逼近克拉美罗界的最小值。以一维可移动天线阵列的感知系统为例，接收端收到的回波信号可表示为

$$\boldsymbol{Y}=\beta\boldsymbol{\alpha}(\boldsymbol{x},u)\boldsymbol{s}^{\mathrm{T}}+\boldsymbol{Z} \tag{26}$$

其中 $\beta$ 是复信道衰落系数，$\boldsymbol{x}=[x_1,x_2,\cdots,x_N]^{\mathrm{T}}$ 是 $N$ 个可移动天线阵元的位置向量，$u=\cos\theta$ 是物理到达角 $\theta$ 所对应的空域到达角，$\boldsymbol{\alpha}(\boldsymbol{x},u)=[\exp(\mathrm{j}2\pi x_1u/\lambda),\exp(\mathrm{j}2\pi x_2u/\lambda),\cdots,\exp(\mathrm{j}2\pi x_Nu/\lambda)]^{\mathrm{T}}$ 是由天线位置 $\boldsymbol{x}$ 和空域到达角 $u$ 所决定的 $N$ 维导向矢量；$\boldsymbol{s}=[s_1,s_2,\cdots,s_T]^{\mathrm{T}}$ 是 $T$ 个快照内由感知目标所反射的回波信号，每次快照的反射回波信号能量的均值为 $P$；$\boldsymbol{Z}$ 是 $N\times T$ 维噪声矩阵，每个矩阵元素为独立同分布复高斯随机变量，其均值为 0，方差为 $\sigma^2$。根据此接收回波信号表达式，感知系统对到达角 $u$ 的估计误差的克拉美罗界可表示为[143]

$$\mathrm{CRB}(\boldsymbol{x})=\frac{\sigma^2\lambda^2}{8\pi^2T\cdot P\cdot N\cdot|\beta|^2}\frac{1}{\mathrm{var}(\boldsymbol{x})} \tag{27}$$

其中 $\mathrm{var}(\boldsymbol{x})=\frac{1}{N}\sum_{n=1}^{N}x_n^2-\mu(\boldsymbol{x})^2$ 是向量 $\boldsymbol{x}$ 的方差，$\mu(\boldsymbol{x})=\frac{1}{N}\sum_{n=1}^{N}x_n$ 是向量 $\boldsymbol{x}$ 的均值。上式表明，角度估计误差的克拉美罗界和天线位置 $\boldsymbol{x}$ 有关，即可移动天线的空间位置自由度为提升感知精度带来了额外的自由度。最小化角度估计误差的克拉美罗界的闭式最优天线位置为[143]

$$x_n^*=\begin{cases}(n-1)D, & n=1,2,\cdots,\lfloor N/2\rfloor\\ A-(N-n)D, & n=\lfloor N/2\rfloor+1,\cdots,N\end{cases} \tag{28}$$

其中 $D$ 是最短天线间距，$A$ 是一维区域的长度。

图 19 对比了上述一维可移动天线阵列与传统固定位置天线阵列的角度估计误差及克拉美罗界，其中天线数量均为 20，可移动天线部署在长度为 20 个波长的一维区域，天线最小间隔为半波长，固定天线阵列采用半波长间距均匀阵列或最大孔径均匀阵列（即 20 根天线均匀铺满整个一维区域），角度估计采用 MUSIC 算法。可以发现，可移动天线阵列具有最小的克拉美罗界，并且 MUSIC 算法可以在信噪比超过 5 dB 的情况下达到角度估计误差的克拉美罗界；相比之下，半波长均匀阵列克拉美罗界较大，而最大孔径均匀阵列有较小的克拉美罗界，但由于其阵列稀疏性导致了角度模糊，使得实际的角度估计误差很大，无法达到克拉美罗界。

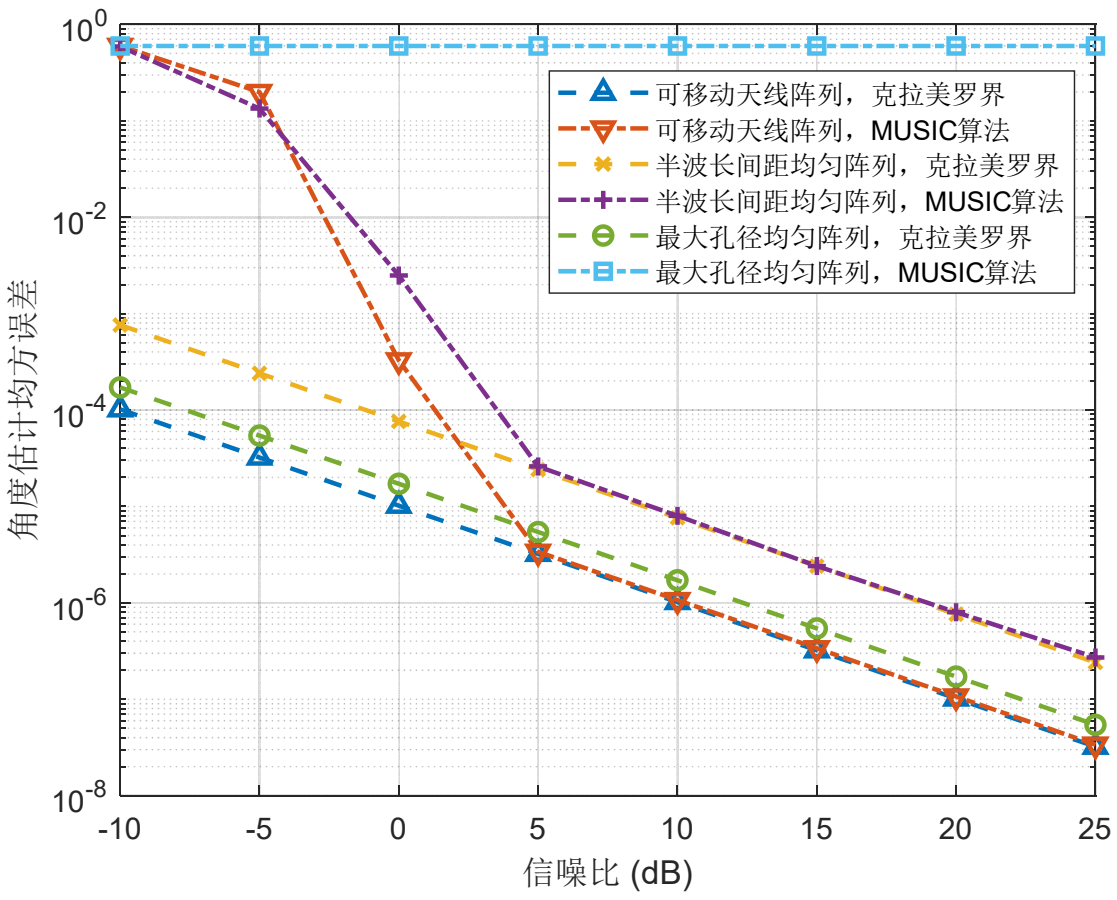

图 19　可移动天线和固定位置天线阵列角度估计均方误差对比

Fig. 19　The comparison of angle estimation error of movable antenna and fixed-position antenna arrays

鉴于可移动天线在通信和感知系统的性能优势，其在通感一体化系统中也具有可观的技术潜力和应用

前景。通过优化天线的位置或朝向等空间自由度，可以显著改善通信及感知性能。然而，由于通信与感知的目标不同，可移动天线的位置/朝向需要合理设计，从而实现通感性能的最佳折中。例如，文献[144]通过联合优化一维可移动天线/流体天线离散位置及发射波束赋形，在满足通信及感知性能需求的前提下，基站发射功率相较于传统固定位置天线可降低 33%。文献[145]进一步研究了二维可移动天线/流体天线辅助的通感一体化系统，通过采用深度强化学习的方法联合优化离散天线位置和预编码，在满足感知需求约束的条件下显著提升多用户通信速率。在此基础上，文献[146]考虑了天线连续位置的优化，在保证最小雷达探测功率的前提下，最大化多用户通信速率。仿真结果表明，在给定天线数量及功率预算的条件下，可移动天线阵列在较小范围内的移动就可以显著提升通感一体化系统性能。此外，文献[147]进一步研究了通感一体化系统中可移动天线阵列和波束赋形的联合设计，在给定总发射功率预算的条件下最大化通信速率于感知互信息的加权和。研究表明，随着天线移动范围的扩大，通感综合性能相较于传统固定天线阵列可提升近 60%。

表 6 总结了可移动天线在无线通信及无线感知领域的主要研究内容及代表性论文。

**表 6　可移动天线天线系统代表性论文**

**Tab. 6　Representative articles in movable antenna systems**

| 研究内容 | | 代表性论文 |
|---|---|---|
| 无线通信/性能分析及优化 | 信道建模 | 场效应信道模型[37-38]，空间相关性信道模型[39-40][111-112] |
| | 空间分集增益 | 信号增强[38-40][118]，干扰抑制[37][48]，RIS/IRS[119-120] |
| | 空间复用增益 | 信道容量提升[110][113][121-127]，发射功率降低[128-133]，安全通信[134-135] |
| | 灵活波束赋形 | 波束调零[135][138]，多波束赋形[137] |
| 无线感知/参数估计 | 信道估计 | 角度域信道估计[114-115]，空间域信道估计[116-117] |
| | 无线感知 | 定位[139-140]，方向估计[140][143]，雷达/成像[141-142] |
| 通感一体化 | 天线位置优化 | 离散位置优化[144-145]，连续位置优化[146-147] |

### 5.3　可移动天线通信与感知小结

随着 6G 网络的通信感知等业务需求的快速增长，传统的固定天线技术依赖于增加天线及射频数量来提升服务质量，导致硬件成本及功耗显著提升。可移动天线技术通过利用空间自由度，在不增加天线及射频数量的前提下可以显著提升通信及感知性能，因此在未来 6G 网络中具有重要的应用潜力。但是，基于机械结构的可移动天线/流体天线的响应速度较慢，主要适用于信道变化速度较慢的场景，例如机器类通信[37]。相比之下，基于电控结构的可移动天线/流体天线，例如像素天线[39]，具有较高的响应速度和较低的功耗，适用于信道快变场景。此外，在实际的移动通信系统中信道呈现快衰落，受限于天线响应速度，很难实现根据瞬时 CSI 及时调整天线位置/朝向。在这种情况下，需要合理设计天线移动的频率及方式，实现通信感知性能与天线移动开销的按需折中。例如，在移动通信系统中，根据用户分布或统计信道信息，以较低的速度和频率调整天线位置/朝向；在通感一体化系统中，根据通信和感知的功能或需求变化，在较长的时间尺度下调整天线位置/朝向。上述实现方法不会大幅改变蜂窝网络的信号处理方式，仅需要在现有架构的基础上增加天线移动控制单元，根据现网协议可测得的数据信息，例如用户接收信号功率、信干比等，调整天线位置或朝向。相应地，随着天线位置/朝向的响应速度降低，通信/感知性能增益也会下降，但相较于传统固定天线仍有潜在的性能优势。除此之外，天线移动会导致天线单元之间耦合效应的改变以及天线辐射增益的变化，应当进一步开展建模和测量工作，以保证其在实际系统中的可靠性。

## 6　总结与展望

### 6.1　总结

本文介绍了面向 6G 通信感知一体化的多天线技术，包括以传统紧凑式阵列和新兴稀疏阵列为代表的集中式天线阵列、分布式天线以及可移动天线/流体天线。为了增加阵列孔径，提升通信抗干扰能力及感知空间分辨率，传统紧凑式天线阵列受制于阵元间距半波长的约束，仅能通过增加阵元数来实现。另一方面，通过打破阵元间距的限制，稀疏天线阵列可以在不增加阵元数的情况下，灵活确定阵元位置，有效提升阵

列总孔径，从而提高空间分辨率，降低硬件及信号处理成本。相比于紧凑式天线阵列，同阵元数目的稀疏天线阵列能形成更窄的主瓣波束，因而具有更高的空间分辨率与干扰抑制能力，然而其产生的栅瓣问题带来角度模糊，为通信与感知系统优化设计带来了新的挑战。不同于均匀稀疏阵列，模块化天线阵列可以通过灵活确定模块数量与每个模块规模，从而实现栅瓣位置和强度的灵活控制，值得进一步深入研究。此外，不同于部署于同一平台的集中式天线阵列，分布式天线阵列将天线布置在不同的地理位置，提供宏分集增益，实现更广域更均匀的覆盖，并通过联合处理有效提升通信与感知性能。然而，面向通信感知一体化的分布式天线架构对同步要求更为严苛、信道硬化效果减弱导致额外的信道估计开销以及前传容量受限等实际挑战。可移动天线或流体天线架构通过在三维立体空间连续优化天线位置，能够获得完全的空间分集增益、空间复用增益以及灵活波束赋形，已引起了国内外学者的极大关注。然而，高效率、低能耗、灵活可控的可移动天线仍需进一步研究。

### 6.2 未来展望

面向 6G 通信感知一体化多天线技术未来主要研究方向主要总结为 4 类，其中包括通感一体化多天线信道状态信息获取新范式、多天线通信感知一体化混合远近场信道建模、通感一体化多天线信息理论分析以及面向智能天线通信感知一体化交叉领域研究，具体如下：

(1) 通感一体化多天线信道状态信息获取新范式：随着天线规模的持续扩展及阵列形态的灵活多变，传统严重依赖导频在线估计的信道状态信息获取模式难以满足快速增长的信道信息需求。近年来提出的信道知识地图[148-149]等方法通过融合区域内所有链路的海量历史数据，构建反映无线信道本质特征的信道知识库，直接根据终端位置信息或虚拟位置信息提前获取部分环境先验信息，从而避免对环境静态要素的重复感知与信道估计，加速甚至避免复杂的在线信道估计，为面向通信感知一体化的多天线系统优化设计提供了新的思路，值得进一步深入研究。

(2) 多天线通信感知一体化混合远近场信道建模：随着天线规模的扩张，用户或散射体可能位于天线阵列的近场区域或者远场区域，并且灵活多变的阵列形态会导致天线单元之间耦合效应的改变以及天线辐射增益的变化。因此如何突破传统基于远场均匀平面波假设及随机统计方法的信道建模局限，建立适应不同天线架构且精准反映非均匀球面波特性以及信道非平稳特性的混合远近场通信感知融合信道模型，是未来的一个重要研究方向。

(3) 通感一体化多天线信息理论分析：当前基于信息理论通感性能分析主要基于传统的半波长集中式天线阵列架构，缺少基于其他 6G 多天线架构通感一体化性能分析。因此如何构建通感一体化多天线性能指标体系，探明不同天线架构下混合远近场通信与感知的性能理论极限及二者的折中与互惠机理，值得进一步深入研究。

(4) 面向智能天线通信感知一体化交叉领域研究：作为一个新兴研究领域，面向智能天线通信感知一体化系统涉及到许多有关通信和感知的交叉领域研究。在信号处理方面，如何设计面向通信感知一体化的波形和信号处理方法是一个亟须解决的问题。其中，面向多天线系统的算法复杂度优化尤为重要。智能天线技术的引入为通信和感知提供了丰富的设计自由度和海量的数据，但如何设计面向智能多天线的通信感知一体化算法是一个关键问题。面对海量的通信和感知数据，人工智能和机器学习技术的发展为面向智能天线的通信感知一体化发展提供了解决途径。因此，如何利用机器学习和人工智能技术优化通感一体算法及系统整体决策，并增强系统的自适应能力是一个关键问题。此外，在网络设计方面，面对未来网络功能多元化和结构多层化的发展趋势，如何进行跨层次通信感知联合优化，也是未来的研究趋势。

### 参考文献

**作者简介**

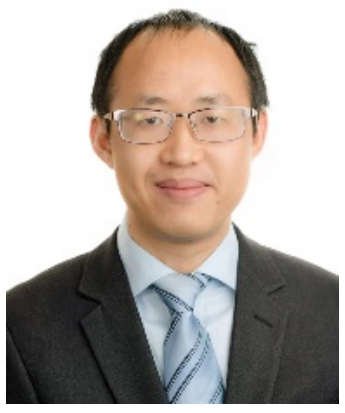

**曾 勇（通讯作者）** 男，1986 年生，湖南邵阳人。东南大学教授，主要研究方向包括无人机通信、超大规模 MIMO、信道知识地图、通感一体化等。
E-mail：yong_zeng@seu.edu.cn

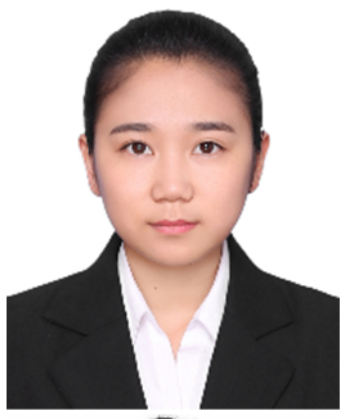

**董珍君** 女，1995 年生，江苏泰州人。东南大学在读博士研究生，主要研究方向包括超大规模 MIMO、信道建模等。
E-mail：zhenjun_dong@seu.edu.cn

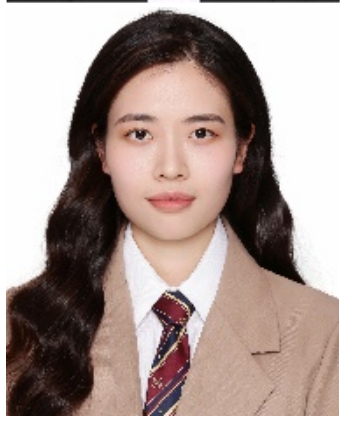

**王蕙质** 女，1999 年生，江苏南京人。东南大学在读硕士研究生，主要研究方向包括超大规模 MIMO、通感一体化等。
E-mail：wanghuizhi@seu.edu.cn

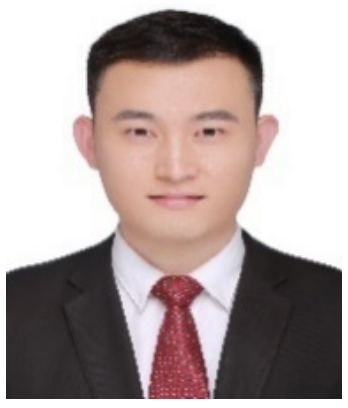

**朱立鹏** 男，1995 年生，河北承德人。新加坡国立大学博士后研究员，主要研究方向为可移动天线辅助的无线通信、智能反射面、无人机通信、毫米波通信、多址接入技术等。
E-mail: zhulp@nus.edu.sg

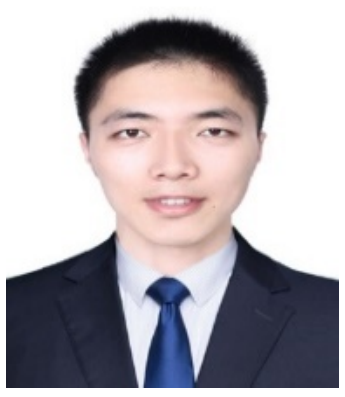

**洪子尧** 男，1996 年生，江苏扬州人。东南大学在读博士研究生，主要研究方向为无蜂窝大规模 MIMO 物理层和分布式计算优化。
E-mail: ziyaohong@seu.edu.cn

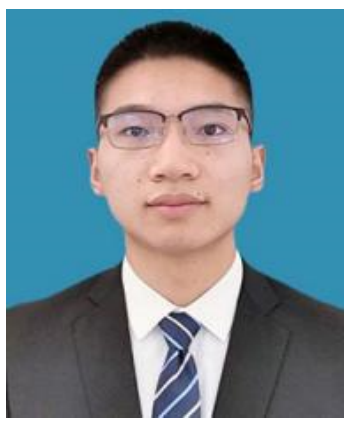

**姜庆基** 男，1998 年生，江苏盐城人。东南大学在读博士研究生，主要研究方向为无蜂窝大规模 MIMO 技术和通感一体化。
E-mail: jiangqingji@seu.edu.cn

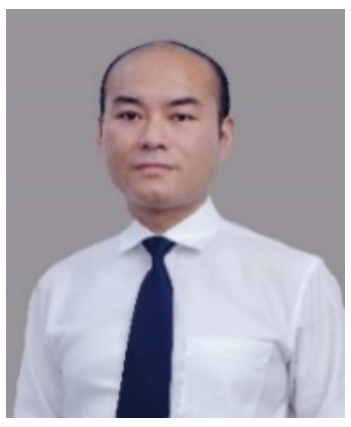

**王东明** 男，1977 年生，河南洛阳人。东南大学教授，主要研究方向为移动通信系统、通信信号处理、多天线传输技术。
E-mail: wangdm@seu.edu.cn

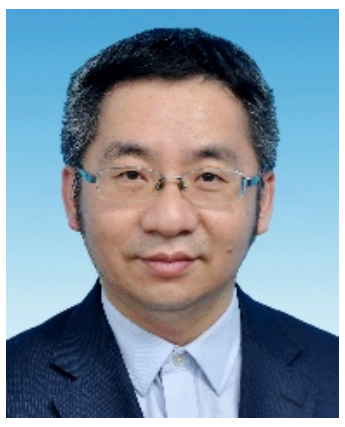

**金 石** 男，1974 年生，安徽黄山人。东南大学移动通信国家重点实验室教授、博士生导师，主要研究方向为 5G/B5G 移动通信理论与关键技术研究、物联网理论与关键技术研究和机器学习与大数据处理在移动通信中的应用等。
E-mail: jinshi@seu.edu.cn

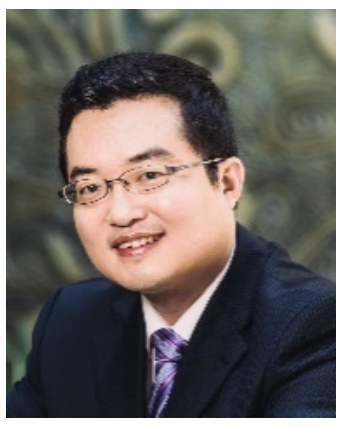

**张 瑞** 男，1976 年生，江苏南京人。香港中文大学（深圳）教授、新加坡国立大学教授，主要研究方向为无人机通信、卫星通信、无线能量传输、智能反射面、可重构MIMO、频谱地图、优化理论等。
E-mail: rzhang@cuhk.edu.cn, elezhang@nus.edu.sg